\documentclass[aps,pra,amssymb,amsmath,eqsecnum,nofootinbib]{revtex4}
\usepackage{graphicx}

\newtheorem{definition}{Definition}[section]
\newtheorem{theorem}{Theorem}[section]
\newtheorem{proposition}{Proposition}[section]
\newtheorem{lemma}{Lemma}[section]
\newtheorem{corollary}{Corollary}[section]
\newtheorem{conjecture}{Conjecture}[section]
\newtheorem{remark}{Remark}[section]

\newenvironment{hypothesis}{HP: \begin{center}} {\end{center}}
\newenvironment{thesis}{TH: \begin{center}} {\end{center}}
\newenvironment{proof}{\begin{center}PROOF: \end{center}} {$ \blacksquare $}
\newtheorem{example}{Example}[section]
\begin{document}
\title{An Introduction to Hyperbolic Analysis}
\author{Andrei Khrennikov  Gavriel Segre}
\email{Andrei.Krennikov@msi.vxu.se, Gavriel.Segre@msi.vxu.se}
 \affiliation{International Center
for Mathematical Modelling in Physics and Cognitive Sciences,
University of V\"{a}xj\"{o}, S-35195, Sweden}
 \maketitle
\newpage
\tableofcontents
\newpage
\section{Introduction}
 Let us consider the ring ${\mathbb{G}} $ of the numbers of the form  $ z \; = \; a + i b \, \, a, b \in
{\mathbb{R}} $, with i satisfying the following equation:
\begin{equation}
    i^{2} \; = \;  1
\end{equation}
The elements of such a ring has been called in the mathematical
literature with different names (cfr. \cite{Jancevic-96} and
references therein): \emph{hyperbolic numbers}, \emph{double
numbers}, \emph{split complex numbers}, \emph{perplex numbers},
and \emph{duplex numbers} \footnote{We invite, by the way, the
reader to pay attention to the fact that, contrary to what is
claimed in \cite{Jancevic-96}, the locution \emph{unipodal
numbers} is used by Garrett Sobczyk \cite{Sobczyk-96a} to denote
the more extended number system $ {\mathbb{U}}$ of the numbers of
the form $ z = a + i b , a ,b \in {\mathbb{C}}  , i^{2} = 1$}.

We will call them \emph{hyperbolic numbers}  and we will refer to
i as to the \emph{hyperbolic imaginary unit}.

Hyperbolic numbers emerged in the research of one of the authors
\cite{Khrennikov-04}, \cite{Khrennikov-03a}, \cite{
Khrennikov-03b} as the underlying number system of a mathematical
theory, "Hyperbolic Quantum Mechanics", axiomatized in the
mentioned papers.

Since, as we will show in section\ref{sec:The hyperbolic algebra
as a bidimensional Clifford algebra}, the complex field
${\mathbb{C}}$ and the hyperbolic ring ${\mathbb{G}}$ are the two
bidimensional Clifford algebras:
\begin{eqnarray}
  {\mathbb{C}} \; &=& \; Cl_{0,1} \\
  {\mathbb{G}} \; &=& \; Cl_{1,0}
\end{eqnarray}
the investigations about Hyperbolic Calculus can be developed
along two different lines:
\begin{enumerate}
    \item one can analyze which of the mirabilities of Complex
    Calculus survives passing to the hyperbolic case
    \item one can directly consider Clifford Calculus
    \cite{Gurlebeck-Sprossig-97}, \cite{Hestenes-Sobczyk-87} in
    its generality and to apply it to the particular (1,0) case
\end{enumerate}
In this paper we will try to pursue both these strategies.

\begin{remark}
\end{remark}
ON OUR USE OF THE LOCUTION "HYPERBOLIC PLANE"

We advise the reader that our adoption of the locution "Hyperbolic
Analysis" follows the terminology of \cite{Khrennikov-99a}.
Consequentially the locution "hyperbolic plane" is used here to
denote $ {\mathbb{G}} $ and has no relation with the more common
use of such a locution to denote the Riemannian manifold $ ( \{
(x,y) \in {\mathbb{R}}^{2} : y > 0 \} \, , \, \frac{ dx \otimes dx
+ dy \otimes dy}{y^{2} } )$

\newpage
\section{The hyperbolic algebra as a bidimensional Clifford
algebra} \label{sec:The hyperbolic algebra as a bidimensional
Clifford algebra}
 Let us consider the ring ${\mathbb{G}} $ of the
hyperbolic numbers, i.e. of the numbers of the form  $ z \; = \; a
+  i b \, \, a, b \in {\mathbb{R}} $, with the hyperbolic
imaginary unit i satisfying the following equation:
\begin{equation}
    i^{2} \; = \; + 1
\end{equation}
$ {\mathbb{G}} $ may be seen as a bidimensional Clifford algebra
as we will show in this paragraph.

Given a linear space V over the real field:
\begin{definition}
\end{definition}
TENSORS OF TYPE (r,s) OVER V:
\begin{equation}
    {\mathcal{T}}^{r}_{s} (V) \; := \; \{ T^{r}_{s} : \times_{i=1}^{r} V \times_{j=1}^{s} V^{\star} \mapsto {\mathbb{R}} \text{ multilinear}  \}
\end{equation}
Let us introduce in particular the following:
\begin{definition}
\end{definition}
TENSOR ALGEBRA OVER V:
\begin{equation}
    T(V) \; := \; \oplus_{ n \in {\mathbb{N}}_{+}}  {\mathcal{T}}^{n}_{0} (V)
\end{equation}

Given an (r,s)-tensor $ T^{r}_{s} \in {\mathcal{T}}^{r}_{s} (V) $
over V:
\begin{definition}
\end{definition}
$ T^{r}_{s} $ IS SKEW-SYMMETRIC IN THE INDICES $ (i,j) $:
\begin{equation}
    T^{r}_{s} \mapsto - T^{r}_{s} \text{ under permutation of i and j}
\end{equation}
\begin{definition}
\end{definition}
$ T^{r}_{s} $ IS SKEW-SYMMETRIC
\begin{equation}
 T^{r}_{s} \text{ skew-symmetric in (i,j) } \forall (i,j)
\end{equation}
\begin{definition}
\end{definition}
EXTERIOR ALGEBRA OVER V:
\begin{equation}
  \wedge^{\star} V \; := \; ( \cup_{p \in {\mathbb{N}}} \wedge^{p} V
   \, , \, \wedge )
\end{equation}
where:
\begin{equation}
  \wedge^{p} V \; := \;  \{ T^{p}_{0} \in {\mathcal{T}}^{p}_{0} (V) \, : \, T^{p}_{0} \text{ skew-symmetric}  \}
\end{equation}
with:
\begin{equation}
 A \wedge B \; := \;  {\mathbb{A}} ( A
 \otimes B) \in \wedge^{r+s} V \; \; A \in \wedge^{r} V , B \in  \wedge^{s} V
\end{equation}
\begin{equation}
     {\mathbb{A}} (T) ( v^{1} , \cdots , v^{r} )  \; := \; \frac{1}{r !} \sum_{ p \in S_{r} }
     sign(p) T( v^{p(1)} , \cdots , v^{p(r)} ) \; \; T \in
     {\mathcal{T}}^{r}_{0} (V) , v_{1} , \cdots , v_{r} \in V
\end{equation}
Given a scalar product $ q : V \times V \mapsto {\mathbb{R}} $
over V:
\begin{definition}
\end{definition}
\begin{equation}
  {\mathcal{I}}_{q} \; := \; \{ x \otimes v \otimes v + q(v,v) \otimes y  \; \; x,y \in T( V) , v \in V \}
\end{equation}
One has that \cite{De-Bartolomeis-93}:
\begin{theorem} \label{th:direct sum decomposition of the tensor algebra induced by a scalar product}
\end{theorem}
\begin{equation}
    T(V) \; = \; \wedge^{\star} V  \, \oplus \, {\mathcal{I}}_{q}
\end{equation}
Denoted by $ \pi_{q} : T(V) \mapsto \wedge^{\star} V $ the
projection induced by the direct sum decomposition of
theorem\ref{th:direct sum decomposition of the tensor algebra
induced by a scalar product} let us introduce the following:
\begin{definition}
\end{definition}
CLIFFORD ALGEBRA ON V W.R.T. q:

the algebra $ Cl_{q}(V) \, := \, ( \wedge^{\star} V , \cdot ) $:
\begin{equation}
    \alpha \cdot \beta \; := \; \pi_{q} ( s \otimes t)  \; \; s \in \pi_{q}^{- 1} ( \alpha
    ),t \in \pi_{q}^{- 1} ( \beta )
\end{equation}
Let us recall that, given a scalar product q on V, one can
introduce the following:
\begin{definition}
\end{definition}
QUADRATIC FORM W.R.T. q:

the map $ Q : V \mapsto {\mathbb{R}} $:
\begin{equation}
    Q(v) \; := \; q( v , v)
\end{equation}
Let us recall, furthermore, the following:
\begin{definition}
\end{definition}
q IS NONDEGENERATE:
\begin{equation}
    q(v,w) = 0 \; \; \forall v \in V \; \Rightarrow \; w = 0
\end{equation}
We can at last present the following:
\begin{theorem}
\end{theorem}
SYLVESTER'S THEOREM:

\begin{hypothesis}
\end{hypothesis}
\begin{equation}
    dim_{{\mathbb{R}}} V \; = \; n
\end{equation}

\begin{equation}
    B \; := \; \{ e_{1} , \cdots , e_{n} \} \text{ basis of V}
\end{equation}
\begin{equation}
    q : V \times V \mapsto {\mathbb{R}} \text{ scalar product :}
    \;
     Q(x) \, = \; \sum_{i=1}^{p} x_{i}^{2} - \sum_{i=1}^{r}
     x_{i}^{2} \, \, x = \sum_{i=1}^{n} x_{i} e_{i} \, \, p+r = n
\end{equation}
\begin{thesis}
\end{thesis}
\begin{equation}
    sign(V,q) \; := \; (p,r)  \text{ is B-independent}
\end{equation}
\begin{definition} \label{def:Clifford algebra with signature (p,r)}
\end{definition}
\begin{equation}
    Cl_{p,r} \; := \; Cl_{q}( {\mathbb{R}}^{p+r} ) \, : \; sign(
    {\mathbb{R}}^{p+r}, q) \, = \, (p,r)
\end{equation}
\begin{definition}
\end{definition}
PAULI MATRICES:
\begin{eqnarray}
  \sigma_{0} \; :&=& \; \left(%
\begin{array}{cc}
  1 & 0 \\
  0 & 1 \\
\end{array}%
\right) \\
  \sigma_{1} \; :&=& \; \left(%
\begin{array}{cc}
  0 & 1 \\
  1 & 0 \\
\end{array}%
\right) \\
  \sigma_{2} \; :&=& \left(%
\begin{array}{cc}
  0 & -i \\
  i & 0 \\
\end{array}%
\right)   \\
 \sigma_{3} \;  :&=& \; \left(%
\begin{array}{cc}
  1 & 0 \\
  0 & -1 \\
\end{array}%
\right)
\end{eqnarray}
\begin{theorem}\label{th:classification of bidimensional Clifford algebras}
\end{theorem}
CLASSIFICATION OF BIDIMENSIONAL CLIFFORD ALGEBRAS
\begin{enumerate}
    \item
\begin{equation}
    Cl_{1,0} \; = \;  {\mathbb{G}} \; = \; {\mathbb{R}} \oplus {\mathbb{R}}
\end{equation}
    \item
\begin{equation}
   Cl_{0,1} \; = \; {\mathbb{C}}
\end{equation}
\end{enumerate}
\begin{proof}
\begin{enumerate}
    \item
Introduced the algebra:
\begin{equation}
    S_{1,0} \; := \; \{ x := x_{1} \sigma_{0} + x_{2} \sigma_{1}
    \;  x_{1} , x_{2} \in {\mathbb{R}} \}
\end{equation}
(with sum and product given, respectively, by matricial sum and
matricial multiplication) and the linear map $ Is_{1,0} :
{\mathbb{R}}^{2} \mapsto  S_{1,0} $:
\begin{equation}
  Is_{1,0}  ( \left(%
\begin{array}{c}
  x_{1} \\
  x_{2} \\
\end{array}%
\right)) \; := \; x_{1} \sigma_{0} + x_{2} \sigma_{1} \; = \; \left(%
\begin{array}{cc}
  x_{1} & x_{2} \\
  x_{2} & x_{1} \\
\end{array}%
\right)
\end{equation}
one has that:
\begin{equation}
    Is_{1,0} ( \left(%
\begin{array}{c}
  x_{1} \\
  x_{2} \\
\end{array}%
\right)) \cdot Is_{1,0} ( \left(%
\begin{array}{c}
  y_{1} \\
  y_{2} \\
\end{array}%
\right)) \; = \; \left(%
\begin{array}{cc}
  x_{1} & x_{2} \\
  x_{2} & x_{1} \\
\end{array}%
\right) \cdot \left(%
\begin{array}{cc}
  y_{1} & y_{2} \\
  y_{2} & y_{1} \\
\end{array}%
\right) \; = \; \left(%
\begin{array}{cc}
  x_{1} y_{1} + x_{2} y_{2} &  x_{1} y_{2} + x_{2} y_{1} \\
  x_{1} y_{2} + x_{2} y_{1} &  x_{1} y_{1} + x_{2} y_{2} \\
\end{array}%
\right)
\end{equation}
is an algebra isomorphism among $ ( S_{1,0} , + , \cdot )$ and $ (
{\mathbb{R}} \oplus {\mathbb{R}}, + ,\cdot_{1,0} )$ where in the
latter algebra the sum is the usual sum of vectors in $
{\mathbb{R}}^{2} $ while the product is given by:
\begin{equation}
    \left(%
\begin{array}{c}
  x_{1} \\
  x_{2} \\
\end{array}%
\right) \, \cdot_{1,0} \, \left(%
\begin{array}{c}
  y_{1} \\
  y_{2} \\
\end{array}%
\right) \; = \; \left(%
\begin{array}{c}
  x_{1} y_{1} + x_{2} y_{2} \\
  x_{1} y_{2} + x_{2} y_{1} \\
\end{array}%
\right)
\end{equation}
From the other side the isomorphism among $ ( {\mathbb{R}} \oplus
{\mathbb{R}}, + ,\cdot_{1,0} ) $ and $ ( {\mathbb{G}} , + , \cdot
) $ appears evident as soon as one makes the identification:
\begin{equation}
    i \;  \equiv \; \left(%
\begin{array}{c}
  0 \\
  1 \\
\end{array}%
\right)
\end{equation}
In particular:
\begin{equation}
  i^{2} \; \equiv \; \left(%
\begin{array}{c}
  0 \\
  1 \\
\end{array}%
\right) \, \cdot_{1,0} \, \left(%
\begin{array}{c}
  0 \\
  1 \\
\end{array}%
\right) \; = \; + 1 \; \equiv \; \sigma_{1} \cdot \sigma_{1}
\end{equation}
    \item
Introduced the algebra:
\begin{equation}
    S_{0,1} \; := \; \{ x := x_{1} \sigma_{0} + x_{2} (i
    \sigma_{2})
    \;  x_{1} , x_{2} \in {\mathbb{R}} \}
\end{equation}
where i is the usual complex imaginary unit such that $ i^{2} \; =
\; -1 $ (with sum and product given,respectively, by matricial sum
and matricial multiplication) and the linear map $ Is_{0,1} :
{\mathbb{R}}^{2} \mapsto  S_{0,1} $:
\begin{equation}
  Is_{0,1}  ( \left(%
\begin{array}{c}
  x_{1} \\
  x_{2} \\
\end{array}%
\right)) \; := \; x_{1} \sigma_{0} + x_{2} ( i \sigma_{2}) \; = \; \left(%
\begin{array}{cc}
  x_{1} & x_{2} \\
  - x_{2} & x_{1} \\
\end{array}%
\right)
\end{equation}
one has that:
\begin{equation}
    Is_{0,1} ( \left(%
\begin{array}{c}
  x_{1} \\
  x_{2} \\
\end{array}%
\right)) \cdot Is_{0,1} ( \left(%
\begin{array}{c}
  y_{1} \\
  y_{2} \\
\end{array}%
\right))\; = \; \left(%
\begin{array}{cc}
  x_{1} & x_{2} \\
  - x_{2} & x_{1} \\
\end{array}%
\right) \cdot \left(%
\begin{array}{cc}
  y_{1} & y_{2} \\
  - y_{2} & y_{1} \\
\end{array}%
\right) \; = \; \left(%
\begin{array}{cc}
  x_{1} y_{1} - x_{2} y_{2} &  x_{1} y_{2} + x_{2} y_{1} \\
  - x_{1} y_{2} -x_{2} y_{1} &  x_{1} y_{1} - x_{2} y_{2} \\
\end{array}%
\right)
\end{equation}
is an algebra isomorphism among $ ( S_{0,1} , + , \cdot )$ and $ (
{\mathbb{R}}^{2}, + ,\cdot_{0,1} )$ where in the latter algebra
the sum is the usual sum of vectors in $ {\mathbb{R}}^{2} $ while
the product is given by;
\begin{equation}
    \left(%
\begin{array}{c}
  x_{1} \\
  x_{2} \\
\end{array}%
\right) \, \cdot_{0,1} \, \left(%
\begin{array}{c}
  y_{1} \\
  y_{2} \\
\end{array}%
\right) \; = \; \left(%
\begin{array}{c}
  x_{1} y_{1} - x_{2} y_{2} \\
  x_{1} y_{2} + x_{2} y_{1} \\
\end{array}%
\right)
\end{equation}
From the other side the isomorphism among $ ( {\mathbb{R}}^{2}  +
, \cdot_{0,1} ) $ and $ ( {\mathbb{C}} , + , \cdot ) $ appears
evident as soon as one makes the identification:
\begin{equation}
    i \;  \equiv \; \left(%
\begin{array}{c}
  0 \\
  1 \\
\end{array}%
\right)
\end{equation}
In particular:
\begin{equation}
  i^{2} \; \equiv \; \left(%
\begin{array}{c}
  0 \\
  1 \\
\end{array}%
\right) \, \cdot_{0,1} \, \left(%
\begin{array}{c}
  0 \\
  1 \\
\end{array}%
\right) \; = \; - 1 \; \equiv \; (i \sigma_{2}) \cdot ( i
\sigma_{2} )
\end{equation}
\end{enumerate}

\end{proof}

According to theorem\ref{th:classification of bidimensional
Clifford algebras} both Complex Analysis and Hyperbolic Analysis
may be seen as Analysis over a bidimensional Clifford algebra, the
passage from the former to the latter corresponding to the ansatz:
\begin{equation}
    sign( {\mathbb{R}}^{2} , q) \, = \, (0,1) \; \mapsto \;  sign( {\mathbb{R}}^{2} , q) \, = \, (1,0)
\end{equation}

\newpage
\section{Limits and series in the hyperbolic plane}
Let us consider the ring ${\mathbb{G}} $ of the hyperbolic
numbers, i.e. of the numbers of the form  $ z \; = \; a +  i b \,
\, a, b \in {\mathbb{R}} $, with the hyperbolic imaginary unit i
satisfying the following equation:
\begin{equation}
    i^{2} \; = \; + 1
\end{equation}
\begin{definition}
\end{definition}
HYPERBOLIC CONJUGATE OF $ z \; = \; a + b i \, \, a, b \in
{\mathbb{R}} $
\begin{equation}
 z^{\star} \; := \; a \, - \, i b
\end{equation}
 Given $ z \; = \; a + b i \; \in \; {\mathbb{G}} $:
\begin{definition}
\end{definition}
\begin{equation}
  {\mathbb{G}}_{+} \; := \; \{ z \in {\mathbb{G}} \, : \, | z | := \sqrt{ z  z^{\star}  } \geq 0 \}
\end{equation}
\begin{definition}
\end{definition}
\begin{equation}
  {\mathbb{G}}_{+}^{\star} \; := \; \{ z \in {\mathbb{G}} \, : \, | z | := \sqrt{ z  z^{\star}  } > 0 \}
\end{equation}
Let us now introduce on $  {\mathbb{G}} $ the following norm:
\begin{definition} \label{def:norm of an hyperbolic number}
\end{definition}
\begin{equation}
    \| x + i y \| \; := \; \sqrt{ x^{2} + y^{2} } \; \; x , y \in {\mathbb{R}}
\end{equation}
and the associated metric:
\begin{definition}
\end{definition}
\begin{equation}
    d_{\| \cdot \|} ( z_{1} , z_{2} ) \; := \; \| z_{1} - z_{2} \|
\end{equation}

\begin{remark}
\end{remark}
THE NONISOMORPHISM OF THE ALGEBRAIC STRUCTURES $ ( {\mathbb{G}} ,
+ , \cdot , \| \cdot \| ) $ AND $  ( {\mathbb{C}} ,
+_{{\mathbb{C}}} , \cdot_{{\mathbb{C}}} , \| \cdot \| ) $

One could suspect that the adoption on $  {\mathbb{G}} $ of the
norm of definition\ref{def:norm of an hyperbolic number} results
in the collapse to the complex case.

That this is not the case, anyway, may be immediately realized as
soon as one introduces the complex algebraic structure $  (
{\mathbb{C}} , +_{{\mathbb{C}}} , \cdot_{{\mathbb{C}}} , \| \cdot
\| ) $ and compares it with the hyperbolic one $ ( {\mathbb{G}} ,
+ , \cdot , \| \cdot \| ) $ by introducing the following:
\begin{definition} \label{def:first kind complex transform}
\end{definition}
$ 1^{TH} $ KIND COMPLEX TRANSFORM

the map  $ T : {\mathbb{G}} \mapsto {\mathbb{C}} $:
\begin{equation}
    T ( x_{1} + i  x_{2} ) \; := \; x_{1} + j x_{2}
    \; \; \forall x_{1} , x_{2} \in {\mathbb{R}}
\end{equation}
where $ j \in {\mathbb{C}} $ is the complex imaginary unit such
that:
\begin{equation}
    j^{2} \; := \; j \cdot_{{\mathbb{C}}} j \; = \; - 1
\end{equation}
One has of course that:
\begin{theorem} \label{th:first kind transform preserves the usual absolute value and sums}
\end{theorem}
\begin{enumerate}
    \item
\begin{equation}
    |  T(z) | \; = \; \| z \| \; \; \forall z \in {\mathbb{G}}
\end{equation}
    \item
\begin{equation}
    | T( z_{1} + z_{2} ) | \; = \; \| z_{1} + z_{2} \| \; \; \forall z_{1},z_{2} \in {\mathbb{G}}
\end{equation}
\end{enumerate}

\smallskip
Let us observe anyway that:
\begin{equation}
  | T( z_{1} \cdot z_{2} ) | \; \neq \; \| z_{1} \cdot z_{2} \|
\end{equation}

\smallskip

The  $ d_{\| \cdot \|}$  allows to define limits on $ {\mathbb{G}}
$ in the usual way \cite{Reed-Simon-80}.

Given a map $ f : {\mathbb{G}} \mapsto {\mathbb{G}} $ and two
points $ z_{1} , z_{2} \in {\mathbb{G}} $:
\begin{definition}
\end{definition}
f TENDS TO $ z_{2} $ IN THE LIMIT $ z \rightarrow z_{1} \; \; (
\lim_{z \rightarrow z_{1}} f(z) \; = \; z_{2}) $:
\begin{equation}
    \forall \epsilon > 0 \, , \, \exists \delta > 0 \; : \; d_{\| \cdot
    \|} ( z , z_{1} ) < \delta \; \Rightarrow \; d_{ \| \cdot
    \|} (  f(z) , z_{2} ) < \epsilon
\end{equation}

Given a sequence $ \{ a_{n} \}_{ n \in {\mathbb{N}} } $ of
hyperbolic numbers and a point $ z \in {\mathbb{G}} $ in the
hyperbolic plane:
\begin{definition}
\end{definition}
$ a_{n} $ TENDS TO z IN THE LIMIT $ n \rightarrow \infty \; \; (
\lim_{n \rightarrow \infty} a_{n} \; = \; z) $
\begin{equation}
  \forall \epsilon > 0 \, , \, \exists N \in {\mathbb{N}}  : n > N
  \; \Rightarrow \; d_{ \| \cdot
    \|} (  a_{n} , z ) < \epsilon
\end{equation}
\begin{definition}
\end{definition}
\begin{equation}
    \sum_{n=0}^{\infty} a_{n} z^{n} \; := \; \lim_{n \rightarrow
    \infty} \sum_{i=0}^{n} a_{i} z^{i}
\end{equation}
(of course provided the limit  exists).
\newpage
\section{The hyperbolic Euler formula}
Let us introduce the following maps defined in the points of the
hyperbolic plane where the series converges:
\begin{definition}
\end{definition}
\begin{equation}
    \exp (z ) \; := \; \sum_{n=0}^{\infty} \frac{ z^{n} }{n ! }
\end{equation}
\begin{definition}
\end{definition}
\begin{equation}
    \cosh(z ) \; := \; \sum_{n=0}^{\infty} \frac{ z^{2n} }{(2 n) ! }
\end{equation}
\begin{definition}
\end{definition}
\begin{equation}
    \sinh(z ) \; := \; \sum_{n=0}^{\infty} \frac{ z^{2n+1} }{(2n+1) ! }
\end{equation}
One has that:
\begin{theorem}
\end{theorem}
HYPERBOLIC EULER FORMULA:
\begin{equation}
 \exp ( i \theta ) \; = \;  \cosh(\theta) \, + i \, \sinh(\theta )
 \; \; \forall \theta \in {\mathbb{R}}
\end{equation}
\begin{proof}
\begin{equation}
 \exp ( i \theta )  \; = \; \sum_{n=0}^{\infty} \frac{ ( i \theta )^{2n} }{ ( 2 n )!
 } \, + \, \sum_{n=0}^{\infty} \frac{ ( i \theta )^{2n+1} }{  (2 n
 +1)! }
\end{equation}
Since:
\begin{equation}
    (i)^{2n} \; = \; 1 \; \;  \forall n \in  {\mathbb{N}}
\end{equation}
\begin{equation}
  (i)^{2n+1} \; = \; i \; \;  \forall n \in  {\mathbb{N}}
\end{equation}
the thesis immediately follows
\end{proof}
\newpage
\section{Analytic functions  in the hyperbolic plane}
Given a map $ f : {\mathbb{G}} \rightarrow {\mathbb{G}} $ and a
point $ z_{0} \in {\mathbb{G}}$ of the hyperbolic plane:
\begin{definition} \label{def:derivative of a map}
\end{definition}
DERIVATIVE OF f IN $ z_{0} $:
\begin{equation}
    f' ( z_{0} ) \; := \; \lim_{z \rightarrow z_{0}} \frac{ f(z) -f( z_{0}) }{z - z_{0}}
\end{equation}
\begin{theorem} \label{th:Cauchy-Riemann conditions in the hyperbolic plane}
\end{theorem}
CAUCHY-RIEMANN CONDITIONS IN THE HYPERBOLIC PLANE:

\begin{hypothesis}
\end{hypothesis}
\begin{equation*}
    z_{0} = x_{0} + i y_{0} \in {\mathbb{G}}
\end{equation*}
\begin{equation*}
   f : {\mathbb{G}} \rightarrow {\mathbb{G}} \; : \exists  f' ( z_{0} )
\end{equation*}
\begin{equation}
    f( x + i y) \; = \; u( x,y) + i v(x,y)
\end{equation}

\begin{thesis}
\end{thesis}

\begin{eqnarray*}
  \frac{ \partial u}{ \partial x} ( x_{0} , y_{0} ) \; &=& \;  \frac{ \partial v}{ \partial y} ( x_{0} , y_{0} )  \\
  \frac{ \partial u}{ \partial y} ( x_{0} , y_{0} ) \; &=& \; \frac{ \partial v}{ \partial x} ( x_{0} , y_{0} )
\end{eqnarray*}
\begin{proof}
 By definition\ref{def:derivative of a map} we have that:
 \begin{equation}
  f' ( z_{0}) \; = \; \lim_{\Delta x \rightarrow 0 ,\Delta y \rightarrow 0
  } ( \frac{ u( x_{0} + i \Delta x_{0}, y_{0} + i \Delta y_{0} )-u( x_{0} , y_{0} )  }{ \Delta x + i \Delta y
  }+ i \frac{ v( x_{0} + i \Delta x_{0}, y_{0} + i \Delta y_{0} )-v( x_{0} , y_{0} )  }{ \Delta x + i \Delta y
  })
\end{equation}
Since this limit has to be always the same for all the paths going
to $ z_{0} $ it has in particular to exist for the two particular
paths in which, respectively, $ \Delta x \; = \; 0 $ and $ \Delta
y \; = \; 0 $.For the first paths we get:
\begin{equation} \label{eq:derivative following the first path}
 f' ( z_{0}) \; = \; \lim_{ \Delta x \rightarrow 0 } \frac{ u (x_{0} + \Delta x_{0} , y_{0})-u(x_{0}, y_{0})  }{ \Delta x
 } \, + \, i \lim_{ \Delta x \rightarrow 0 } \frac{ v (x_{0} + \Delta x_{0} , y_{0})-v(x_{0}, y_{0})}{ \Delta x
 } \; = \; \frac{\partial u }{\partial x }( x_{0},y_{0} ) \, + i \frac{\partial v }{\partial x }( x_{0},y_{0} )
\end{equation}
while for the second path we obtain:
\begin{equation} \label{eq:derivative following the second path}
    f' ( z_{0}) \; = \; \lim_{ \Delta y \rightarrow 0 } \frac{ u (x_{0}  , y_{0}+ \Delta x_{0})-u(x_{0}, y_{0})  }{i  \Delta
    y
 } \, + \, i \lim_{ \Delta y \rightarrow 0 } \frac{ v (x_{0} , y_{0} + \Delta y_{0})-v(x_{0}, y_{0}) }{ i \Delta
 y
 } \; = \; i  \frac{\partial u }{\partial y }( x_{0},y_{0} ) \, +  \frac{\partial v }{\partial y }( x_{0},y_{0} )
\end{equation}
where in the last passage we have used the fact that in the
hyperbolic plane:
\begin{equation}
    \frac{1}{i} \; = \; i
\end{equation}
Equating the real and imaginary part of eq.\ref{eq:derivative
following the first path} and eq.\ref{eq:derivative following the
second path} the thesis follows
\end{proof}
\begin{corollary} \label{cor:expression of the derivative}
\end{corollary}

\begin{hypothesis}
\end{hypothesis}
\begin{equation*}
    z_{0} = x_{0} + i y_{0} \in {\mathbb{G}}
\end{equation*}
\begin{equation*}
   f : {\mathbb{G}} \rightarrow {\mathbb{G}} \; : \exists  f' ( z_{0} )
\end{equation*}
\begin{equation}
    f( x + i y) \; = \; u(x,y) + i v(x,y)
\end{equation}

\begin{thesis}
\end{thesis}
\begin{equation}
 f' ( z_{0}) \; = \; \frac{\partial u }{\partial x }( x_{0},y_{0}
 ) \, + i \frac{\partial v }{\partial x }( x_{0},y_{0} ) \; = \; \frac{\partial v }{\partial y }( x_{0},y_{0}
 ) \, + \, i \frac{\partial u }{\partial y }( x_{0},y_{0} )
\end{equation}

\begin{corollary} \label{cor:real and imaginary part obey the wave equation}
\end{corollary}

\begin{hypothesis}
\end{hypothesis}
\begin{equation*}
    z_{0} = x_{0} + i y_{0} \in {\mathbb{G}}
\end{equation*}
\begin{equation*}
   f : {\mathbb{G}} \rightarrow {\mathbb{G}} \; : \exists  f' ( z_{0} )
\end{equation*}
\begin{equation}
    f( x + i y) \; = \; u(x,y) + i v(x,y)
\end{equation}

\begin{thesis}
\end{thesis}
\begin{enumerate}
    \item
    \begin{equation}
 (\frac{ \partial^{2 }  }{ \partial x^{2} } \, - \,  \frac{ \partial^{2 }  }{ \partial y^{2}
 }) u ( x_{0}, y_{0} ) \; = \; 0
\end{equation}
    \item
    \begin{equation}
 (\frac{ \partial^{2 }  }{ \partial x^{2} } \, - \,  \frac{ \partial^{2 }  }{ \partial y^{2}
 }) v ( x_{0}, y_{0} ) \; = \; 0
\end{equation}
\end{enumerate}

\begin{proof}
\begin{enumerate}
    \item
    Differentiating the first hyperbolic Cauchy-Riemann equation
w.r.t. x one obtains:
\begin{equation} \label{eq:first ausiliary}
    \frac{ \partial^{2 }  }{ \partial x^{2} }  u ( x_{0}, y_{0} )
    \; = \; \frac{\partial}{ \partial x }  \frac{\partial}{ \partial y
    } v( x_{0}, y_{0} )
\end{equation}
Differentiating the second hyperbolic Cauchy-Riemann equation
w.r.t. y one obtains:
\begin{equation}  \label{eq:second ausiliary}
    \frac{ \partial^{2 }  }{ \partial y^{2} }  u ( x_{0}, y_{0} )
    \; = \; \frac{\partial}{ \partial x }  \frac{\partial}{ \partial y
    } v( x_{0}, y_{0} )
\end{equation}
Subtracting the eq.\ref{eq:second ausiliary} from the
eq.\ref{eq:first ausiliary} one obtains the thesis
    \item
    Differentiating the first hyperbolic Cauchy-Riemann equation
w.r.t. y one obtains:
\begin{equation} \label{eq:third ausiliary}
   \partial_{y} \partial_{x} u ( x_{0}, y_{0} ) \; = \;
   \partial_{y}^{2} v ( x_{0}, y_{0} )
\end{equation}
Differentiating the second hyperbolic Cauchy-Riemann equation
w.r.t. x one obtains:
\begin{equation}  \label{eq:fourth ausiliary}
 \partial_{x} \partial_{y} u ( x_{0}, y_{0} ) \; = \;
   \partial_{x}^{2} v ( x_{0}, y_{0} )
\end{equation}
Subtracting the eq.\ref{eq:third ausiliary} from the
eq.\ref{eq:fourth ausiliary} one obtains the thesis
\end{enumerate}
\end{proof}

\smallskip

\begin{definition} \label{def:analycity}
\end{definition}
f IS ANALYTIC IN $ z_{0} \in  {\mathbb{G}} $
\begin{equation}
    \exists O \in {\mathcal{T}}_{{\mathbb{G}}} \; : \; ( z_{0} \in
    O ) \wedge ( \exists  f'(z) \forall z \in O)
\end{equation}
\begin{remark}
\end{remark}
Let us observe that according to definition \ref{def:analycity} it
is not required that $ u , v \in C^{\infty} ( {\mathbb{R}}^{2} ) $
as, in our opinion erroneously, is made in \cite{Motter-Rosa-98}
\begin{example}
\end{example}
\begin{equation}
    f(z) \; := \; z^{2} \; = \; ( x + i y)^{2} \; = \; x^{2} +
    y^{2} + 2 i x y \; = \; u(x,y) + i v( x,y)
\end{equation}
\begin{equation}
     u(x,y) \; = \; x^{2} + y^{2}
\end{equation}
\begin{equation}
    v(x,y) \; = \; 2 x y
\end{equation}
\begin{equation}
   \partial_{x} u \; = 2 x \; = \; \partial_{y} v
\end{equation}
\begin{equation}
    \partial_{y} u \; = 2 y  \; = \;  \partial_{x}  v
\end{equation}
It follows that f(z) is analytic in the whole  hyperbolic plane.
\begin{example}
\end{example}
\begin{equation}
    f( z ) \; := \; \exp (z) \; = \; \exp(x) ( \cosh (y) + i \sinh
    (y) ) \; = \; u(x,y) + i v( x,y)
\end{equation}
\begin{equation}
    u (x,y) \; = \;  \exp (x) \cosh (y)
\end{equation}
\begin{equation}
    v(x,y) \; = \; \exp (x) sinh(y)
\end{equation}
\begin{equation}
    \partial_{x} u \; = \; \exp (x) \cosh (y) \; = \; \partial_{y}
    v
\end{equation}
\begin{equation}
    \partial_{y} u \; = \; \exp (x) sinh(y) \; = \; \partial_{x} v
\end{equation}
It follows that f(z) is analytic in the whole  hyperbolic plane.
\begin{example}
\end{example}
Let us consider the map $ f : {\mathbb{G}} - Diagonals \, \mapsto
\, {{\mathbb{G}}} $:
\begin{equation}
    Diagonals \; := \; \{ x + i y \in {\mathbb{G}} : y = \pm x \}
\end{equation}
\begin{equation}
    f(z) \; := \; \frac{1}{z} \; = \; \frac{x - i y}{x^{2} -
    y^{2}} \; = \; u(x,y) + i v( x,y)
\end{equation}
\begin{equation}
    u(x,y) \; = \; \frac{x}{x^{2}-y^{2}}
\end{equation}
\begin{equation}
    v(x,y) \; = \; \frac{- y}{x^{2}-y^{2}}
\end{equation}
\begin{equation}
 \partial_{x} u \; = \; - \frac{x^{2}+y^{2}}{(x^{2}-y^{2})^{2}} \;
 = \; \partial_{y} v
\end{equation}
\begin{equation}
    \partial_{y} u \; = \; \frac{2 x y}{( x^{2}-y^{2})^{2}} \; =
    \; \partial_{x} v
\end{equation}
It follows that f is analytic in its whole domain of definition $
{{\mathbb{G}}} - Diagonals  $.

\smallskip

Since:
\begin{lemma}
\end{lemma}
\begin{equation}
    z \cdot z^{\star} \; = \; Q_{1,1} ( \left(%
\begin{array}{c}
  x \\
  y \\
\end{array}%
\right)) \; = \; x^{2} - y^{2} \; \; \forall z = x + i y \in
{\mathbb{G}}
\end{equation}
it would appear more natural to look at $  {\mathbb{G}} $ as the
couple $ ( {\mathbb{R}}^{2}, q_{1,1} ) $. So, for prudence, we
will consider all the possible signatures of  the bilinear
symmetric form one is implicitly assuming in the definition of the
notion of angles among two vectors.

So, given  two curves $ \gamma_{1}, \gamma_{2 } $ in  $
{\mathbb{G}} $ intersecting in a point $ z_{0} $ and given $ n,m
\in {\mathbb{N}} \, : \, n+m =2 $:
\begin{definition}
\end{definition}
(n,m)-ANGLE AMONG $ \gamma_{1} $ and $ \gamma_{2}$ IN $ z_{0} $
\begin{equation}
    (n,m)-angle_{z_{0}} ( \gamma_{1}  , \gamma_{2} ) \; := \;
    q_{n,m} ( \hat{e}_{1} , \hat{e}_{2} )
\end{equation}
where $ \hat{e}_{i} $ is the tangent vector to $ \gamma_{i} $ in $
z_{0} \; i=1,2 $.

 Given a function $ f
: {\mathbb{G}} \mapsto  {\mathbb{G}} $:
\begin{definition}
\end{definition}
f IS (n,m)-CONFORMAL in $ z_{0} \in {\mathbb{G}} $
\begin{equation}
    \gamma_{1} ,   \gamma_{2} \text{ curves in $ {\mathbb{G}} $ intersecting in $ z_{0} $ }
    \; \Rightarrow \\
    (n,m)-angle_{f(z_{0})} (  \gamma_{1}'  ,  \gamma_{2}' ) \; = \;
    (n,m)-angle_{z_{0}} ( \gamma_{1} , \gamma_{2})
\end{equation}
where:
\begin{equation}
    \gamma_{i}' \; = \; f( \gamma_{i} ) \; \; i=1,2
\end{equation}

 Contrary to the  analogous situation in Complex Analysis
one has that:
\begin{theorem}
\end{theorem}
\begin{enumerate}
    \item
\begin{equation}
    ( f \; \text{ analytic in $ z_{0} $} \,  \wedge \, f'(z_{0} ) \neq 0 )  \; \nRightarrow \; f \; \text{(2,0)-conformal in $ z_{0} $ }
\end{equation}
    \item
\begin{equation}
    ( f \; \text{ analytic in $ z_{0} $} \,  \wedge \, f'(z_{0} ) \neq 0 )  \; \nRightarrow \; f \; \text{(1,1)-conformal in $ z_{0} $ }
\end{equation}
\end{enumerate}
\begin{proof}
\begin{enumerate}
    \item
    By definition:
\begin{equation}
   (2,0)-angle_{z_{0}} ( \gamma_{1} , \gamma_{2}) \; = \; q_{2,0} ( \hat{e}_{1}
    , \hat{e}_{2} )
\end{equation}
where $ \hat{e}_{i} $ is the unit tangent vector to $ \gamma_{i} $
in $ z_{0} \; i=1,2$.

Similarly:
\begin{equation}
    (2,0)-angle_{f(z_{0})} (  \gamma_{1}'  ,  \gamma_{2}' ) \; = \;
    q_{2,0} ( \hat{e}_{1}' ,  \hat{e}_{2}' )
\end{equation}
where $ \hat{e}_{i}' $ is the unit tangent vector to $ \gamma_{i}'
$ in $ f(z_{0}) \; i=1,2$.
 Since an infinitesimal displacement along $
\gamma_{i} $ may be expressed formally as $ \delta x_{i}
\hat{e}_{x} + \delta y_{i} \hat{e}_{y} $ one has that:
\begin{equation}
 \hat{e}_{i} \; = \; \frac{\delta x_{i}
\hat{e}_{x} + \delta y_{i} \hat{e}_{y}}{\sqrt{(\delta x_{i})^{2}
+(\delta y_{i})^{2}}} \; \; i=1,2
\end{equation}
\begin{equation}
    \hat{e}_{i}' \; = \; \frac{\delta x_{i}'
\hat{e}_{x} + \delta y_{i}' \hat{e}_{y}}{\sqrt{(\delta x_{i}')^{2}
+(\delta y_{i}')^{2}}} \; \; i=1,2
\end{equation}
Therefore:
\begin{equation}
 q_{2,0} (\hat{e}_{1} , \hat{e}_{2} ) \; = \; \frac{ \delta x_{1} \delta x_{2} + \delta y_{1} \delta y_{2}}{\prod_{i=1}^{2} \sqrt{ (\delta x_{i})^{2} +  (\delta y_{i})^{2} }  }
\end{equation}
\begin{equation} \label{eq:(2,0)-angle among the image curves}
 q_{2,0 } (\hat{e}_{1}' , \hat{e}_{2}') \; = \; \frac{ \delta x_{1}' \delta x_{2}' + \delta y_{1}' \delta y_{2}'}{\prod_{i=1}^{2} \sqrt{ (\delta x_{i}')^{2} +  (\delta y_{i}')^{2} }  }
\end{equation}
Substituting the relations:
\begin{eqnarray}
  \delta x_{i} ' \;  &=& \; \partial_{x} u \delta x_{i} +  \partial_{y} u \delta y_{i} \; \; i=1,2  \\
   \delta y_{i} ' \;  &=& \; \partial_{x} v \delta x_{i} +  \partial_{y} v \delta y_{i} \; \; i=1,2
\end{eqnarray}
into the eq.\ref{eq:(2,0)-angle among the image curves} one
obtains:
\begin{equation}
  \hat{e}_{1}' \cdot \hat{e}_{2}' \; = \; \frac{ [ ( \partial_{x} u)^{2} +  (\partial_{x} v)^{2}  ] \delta x_{1} \delta x_{2} + [ ( \partial_{y} u)^{2} +  (\partial_{y} v)^{2}  ] \delta y_{1} \delta y_{2} + ( \partial_{x} u \partial_{y} u + \partial_{x} v \partial_{y} v ) ( \delta x_{1} \delta y_{2}+ \delta x_{2} \delta y_{1} )    }{\prod_{i=1}^{2} \sqrt{[ ( \partial_{x} u)^{2} +  (\partial_{x} v)^{2}  ] ( \delta x_{i})^{2} + [ ( \partial_{y} u)^{2} +  (\partial_{y} v)^{2}  ] ( \delta y_{i})^{2} + 2 ( \partial_{x} u \partial_{y} u + \partial_{x} v \partial_{y} v ) \delta x_{i} \delta y_{i}  } }
\end{equation}
that using the hyperbolic Cauchy-Riemann equations reduces to:
\begin{equation}
 \hat{e}_{1}' \cdot \hat{e}_{2}' \; = \; \frac{ (\vec{\nabla} u)^{2} (\delta x_{1} \delta x_{2} + \delta y_{1} \delta y_{2} ) + ( \vec{\nabla} u \cdot  \vec{\nabla} v    ) (\delta x_{1} \delta y_{2}+ \delta x_{2} \delta y_{1} ) }{\sqrt{\prod_{i=1}^{2} (\vec{\nabla} u)^{2}[( \delta x_{i})^{2}+ (\delta y_{i})^{2}] + 2 \vec{\nabla} u \cdot \vec{\nabla} v   \delta x_{i} \delta y_{i}
 }}  \; \neq \; \frac{ \delta x_{1} \delta x_{2} + \delta y_{1} \delta y_{2}}{\prod_{i=1}^{2} \sqrt{ (\delta x_{i})^{2} +  (\delta y_{i})^{2} }}
\end{equation}
    \item Since an infinitesimal displacement along $
\gamma_{i} $ may be again expressed formally as $ \delta x_{i}
\hat{e}_{x} + \delta y_{i} \hat{e}_{y} $ one has that again:
\begin{equation}
 \hat{e}_{i} \; = \; \frac{\delta x_{i}
\hat{e}_{x} + \delta y_{i} \hat{e}_{y}}{\sqrt{(\delta x_{i})^{2}
+(\delta y_{i})^{2}}} \; \; i=1,2
\end{equation}
\begin{equation}
    \hat{e}_{i}' \; = \; \frac{\delta x_{i}'
\hat{e}_{x} + \delta y_{i}' \hat{e}_{y}}{\sqrt{(\delta x_{i}')^{2}
+(\delta y_{i}')^{2}}} \; \; i=1,2
\end{equation}
Let us now observe that:
\begin{equation}
 q_{1,1} (\hat{e}_{1} , \hat{e}_{2} ) \; = \; \frac{ \delta x_{1} \delta x_{2} - \delta y_{1} \delta y_{2}}{\prod_{i=1}^{2} \sqrt{ (\delta x_{i})^{2} -  (\delta y_{i})^{2} }  }
\end{equation}
\begin{equation} \label{eq:(1,1)-angle among the image curves}
 q_{1,1 } (\hat{e}_{1}' , \hat{e}_{2}') \; = \; \frac{ \delta x_{1}' \delta x_{2}' - \delta y_{1}' \delta y_{2}'}{\prod_{i=1}^{2} \sqrt{ (\delta x_{i}')^{2} -  (\delta y_{i}')^{2} }  }
\end{equation}
\end{enumerate}
Substituting the relations:
\begin{eqnarray}
  \delta x_{i} ' \;  &=& \; \partial_{x} u \delta x_{i} +  \partial_{y} u \delta y_{i} \; \; i=1,2  \\
   \delta y_{i} ' \;  &=& \; \partial_{x} v \delta x_{i} +  \partial_{y} v \delta y_{i} \; \; i=1,2
\end{eqnarray}
into the eq.\ref{eq:(1,1)-angle among the image curves} one
obtains:
\begin{equation}
   \hat{e}_{1}' \cdot  \hat{e}_{2}' \; = \; \frac{ (\partial_{x} u \delta x_{1} +  \partial_{y} u \delta y_{1})(  \partial_{x} u \delta x_{2} +  \partial_{y} u \delta y_{2} )- (\partial_{x} v \delta x_{1} +  \partial_{y} v \delta y_{1})(\partial_{x} v \delta x_{2} +  \partial_{y} v \delta y_{2})  }{\sqrt{(\partial_{x} u \delta x_{1} +  \partial_{y} u \delta y_{1})^{2}-(\partial_{x} v \delta x_{1} +  \partial_{y} v \delta y_{1})^{2}} {\sqrt{(\partial_{x} u \delta x_{2} +  \partial_{y} u \delta y_{2})^{2}-(\partial_{x} v \delta x_{2} +  \partial_{y} v \delta y_{2})^{2}} }}
\end{equation}
that after a tedious computation using the hyperbolic
Cauchy-Riemann conditions may be expressed as:
\begin{equation}
   \hat{e}_{1}' \cdot  \hat{e}_{2}' \; = \;  \frac{[( \partial_{x}u)^{2}-(\partial_{y}u)^{2}]\delta x_{1} \delta x_{2} + [( \partial_{x} v)^{2}-(\partial_{y}v)^{2}]\delta y_{1} \delta y_{2}  + (\partial_{x} u \partial_{y} u  - \partial_{x} v \partial_{y} v) (\delta x_{1} \delta y_{2}+ \delta x_{2} \delta y_{1})   }{ \prod_{i=1}^{2} [( \partial_{x}u)^{2}- (\partial_{y}u)^{2}] (  \delta x_{i}^{2} - \delta y_{i}^{2}) + 2 (  \partial_{x} u  \partial_{y} u -  \partial_{x} v  \partial_{y} v)  \delta x_{i}  \delta y_{i}
   } \neq \hat{e}_{1} \cdot  \hat{e}_{2}
\end{equation}
\end{proof}

\smallskip

Given a function $ f :  {{\mathbb{G}}} \rightarrow {{\mathbb{G}}}$
such that:
\begin{equation}
    f(z) \; = \; f( x+ i y) \; = \; u(x,y) \, + \, i v(x,y)
\end{equation}
 and a continous curve C in $ {{\mathbb{G}}} $:
\begin{definition}
\end{definition}
INTEGRAL OF f ALONG C:
\begin{multline}
    \int_{C} f(z) d z \; = \; \int_{C}( u(x,y) \, + \, i v(x,y) )  d ( x+iy) \; :=
    \;  \int_{C} u(x,y) d x  + i^{2} \int_{C} v(x,y) dy + i ( \int_{C} u(x,y) d
    y + \int_{C} v(x,y) dx ) \\
    = \int_{C} u(x,y) d x  +  v(x,y) dy + i  \int_{C} v(x,y) d
    x + u(x,y) dy
 \end{multline}
 One has the following:
\begin{theorem} \label{th:hyperbolic Cauchy-Goursat theorem}
\end{theorem}
HYPERBOLIC CAUCHY-GOURSAT THEOREM:

\begin{hypothesis}
\end{hypothesis}
\begin{equation}
    G  \subset {\mathbb{G}} \; : \; \pi_{1}(G) = \{ {\mathbb{I}} \}
    \; : \; \text{ f is analytic in G}
\end{equation}
\begin{equation}
    C \subset G \text{ closed smoooth curve not self-linking}
\end{equation}
\begin{thesis}
\end{thesis}
\begin{equation}
    \oint_{C} f(z) d z \; = \; 0
\end{equation}
\begin{proof}
Let us consider $ {\mathbb{G}} $ as the xy-plane of $
{\mathbb{R}}^{3} $.

Introduced the vector fields:
\begin{eqnarray}
  \vec{A}_{1} \;&=& \; (u , v , 0) \\
  \vec{A}_{2} \; &=& \; ( v , u , 0)
\end{eqnarray}
one has that:
\begin{equation}
  \oint_{C} f(z) d z \; = \; \oint_{C} \vec{A}_{1} \cdot d \vec{r}
  + i  \oint_{C} \vec{A}_{2} \cdot d \vec{r}
\end{equation}
By Stokes Theorem one has that:
\begin{equation}
  \oint_{C}  \vec{A}_{i} \cdot d \vec{r} \, = \, 0 \;
  \Leftrightarrow \; \vec{\nabla}  \wedge \vec{A}_{i}(x,y)=0 \; \;
  \forall z = x + i y \in \partial^{-1} C \; \; i=1,2
\end{equation}
Since:
\begin{equation}
    \vec{\nabla}  \wedge \vec{A}_{1}  \; = \; ( \frac{\partial v }{\partial x
    } - \frac{\partial u}{\partial y } ) \hat{e}_{z}
\end{equation}
\begin{equation}
    \vec{\nabla}  \wedge \vec{A}_{2}  \; = \; ( \frac{\partial u }{\partial x
    } - \frac{\partial v}{\partial y }) \hat{e}_{z}
\end{equation}
(where $ \{ \hat{e}_{x} := \left(%
\begin{array}{c}
  1 \\
  0 \\
  0 \\
\end{array}%
\right) \, , \,  \hat{e}_{y} := \left(%
\begin{array}{c}
  0 \\
  1 \\
  0 \\
\end{array}%
\right) \, , \, \hat{e}_{z} := \left(%
\begin{array}{c}
  0 \\
  0 \\
  1 \\
\end{array}%
\right)  \} $ is the canonical basis of $ {\mathbb{R}}^{3} $) the
thesis immediately follows from theorem\ref{th:Cauchy-Riemann
conditions in the hyperbolic plane}
\end{proof}

\begin{example}
\end{example}
\begin{equation}
    c_{a,b} : [ 0 , 2 \pi ) \mapsto {\mathbb{G}} \; \; a ,b \in
    {\mathbb{R}}_{+} \; : \;
     c_{a,b} (t) \; := \; a \cos(t)+ i b \sin(t)
\end{equation}
\begin{equation}
    f (z) \; := \; z^{2} \; = \; u(x,y) + i v(x,y)
\end{equation}
\begin{equation}
    u(x,y) \; = \; x^{2} + y^{2}
\end{equation}
\begin{equation}
    v(x,y) \; = \; 2 x y
\end{equation}
\begin{equation}
    \oint_{c_{a,b}} f(z) dz \; = \;  \oint_{c_{a,b}} ( x^{2} +
    y^{2}) dx \, + \, 2 x y dy \: +  i \: \oint_{c_{a,b}} 2 x y dx
    + ( x^{2} +
    y^{2}) dy
\end{equation}
\begin{multline}
    \oint_{c_{a,b}} f(z) dz \; = \;\int_{0}^{2 \pi} [( a^{2} \cos^{2}(t) + b^{2}
    \sin^{2}(t)) \cdot ( - a \sin(t)) + 2 a  \cos (t) b \sin(t) b
    \cos(t) ] dt \, + \\
     i \int_{0}^{2 \pi} ( - 2 a^{2} b \cos (t)
    \sin^{2}(t) + a^{2} b \cos^{3}(t) + b^{3} \sin^{2} (t) \cos
    (t) ) \; = \; 0
\end{multline}
\begin{example}
\end{example}
\begin{equation}
    c_{a,b} : [ 0 , 2 \pi ) \mapsto {\mathbb{G}} \; \; a ,b \in
    {\mathbb{R}}_{+} \; : \;
     c_{a,b} (t) \; := \; a \cos(t)+i b \sin(t)
\end{equation}
\begin{equation}
    f(z) \; := \; \exp(z) \; = u(x,y) + i v(x,y)
\end{equation}
\begin{equation}
    u(x,y) \; = \; \exp(x) \cosh(y)
\end{equation}
\begin{equation}
    v(x,y) \; = \; \exp(x) \sinh (y)
\end{equation}
\begin{equation}
    \oint_{c_{a,b}} f(z) dz \; = \; \oint_{c_{a,b}} \exp(x)
    \cosh(y) dx +  \exp(x) \sinh (y) dy \, + \, i \oint_{c_{a,b}} \exp(x) \sinh
    (y) dx + \exp(x) \cosh(y) dy
\end{equation}
\begin{multline}
     \oint_{c_{a,b}} f(z) dz \; = \; \int_{0}^{2 \pi} [ \exp ( a
     \cos(t)) \cosh(b \sin(t)) ( - a \sin(t)) + \exp ( a \cos (t))
     \sinh( b \sin (t)) ( b \cos(t))] dt \, + \\
      i \int_{0}^{2
     \pi} [ \exp( a \cos(t)) \sinh(b \sin(t)) ( - a \sin (t) ) + \exp( a
     \cos(t))\cosh ( b \sin(t)) (b \cos (t)) ] \; = \; 0
\end{multline}
\begin{example}
\end{example}
\begin{equation}
    c_{a,b} : [ 0 , 2 \pi ) \mapsto {\mathbb{G}} \; \; a:=1 , b:= \frac{1}{2} \;
    : \; c_{a,b} (t) \; := \; 4+ a \cos(t) + i ( 2 + b \sin(t) )
\end{equation}
\begin{equation}
    f(z) \; := \; \frac{1}{z} \; = \; u(x,y) + i v(x,y)
\end{equation}
\begin{equation}
    u(x,y) \; = \; \frac{x}{ x^{2} - y^{2} }
\end{equation}
\begin{equation}
    v(x,y) \; = \; \frac{- y}{ x^{2} - y^{2} }
\end{equation}
\begin{equation}
    \oint_{c_{a,b}} f(z) dz \; = \; \oint_{c_{a,b}}  \frac{x}{ x^{2} - y^{2}
    } dx +  \frac{- y}{ x^{2} - y^{2} } d y \, + \, i
    \oint_{c_{a,b}} \frac{- y}{ x^{2} - y^{2} } dx + \frac{x}{ x^{2} - y^{2}
    } dy
\end{equation}
\begin{multline}
   \oint_{c_{a,b}} f(z) dz \; = \; \int_{0}^{2 \pi} [ \frac{(4 + a \cos (t)) }{(4 + a \cos (t))^{2} - ( 2 + b \sin (t))^{2}
   } ( - a \sin (t) ) - \frac{( 2 + b \sin (t))}{(4 + a \cos (t))^{2} - ( 2 + b \sin
   (t))^{2}} ( b \cos (t) )] dt  \, + \\
    i \int_{0}^{2 \pi}  [ \frac{-( 2 + b \sin
   (t))}{(4 + a \cos (t))^{2} - ( 2 + b \sin (t))^{2}} ( - a \sin
   (t) ) + \frac{(4 + a \cos (t))}{(4 + a \cos (t))^{2} - ( 2 + b \sin
   (t))^{2}} ( b \cos (t))] dt \; = \; 0
\end{multline}

\smallskip
Let us observe that as in the complex case one has that:
\begin{lemma} \label{lem:upper bound to the integral along a finite curve}
\end{lemma}

\begin{hypothesis}
\end{hypothesis}
\begin{equation}
    \gamma \text { curve in $ {\mathbb{G}} $ not self-linking}
\end{equation}
\begin{equation}
    f : {\mathbb{G}} \rightarrow  {\mathbb{G}} \text{ continuous }
    \, : \,( \exists M \in {\mathbb{R}}_{+} : \| f(z) \| \leq M \;
    \forall z \in \gamma )
\end{equation}
\begin{thesis}
\end{thesis}
\begin{equation}
    \| \int_{\gamma} f(z) dz \| \; \leq \; M L_{\gamma}
\end{equation}
where $ L_{\gamma} $ denotes the length of the curve $ \gamma $
\begin{proof}
\begin{multline}
    \| \int_{\gamma} f(z) dz \| \; = \;  \| \lim_{N \rightarrow\infty , \Delta z_{i} \rightarrow 0
    }  \sum_{i=1}^{N} f( z_{i} ) \Delta z_{i} \| \; = \; \lim_{N \rightarrow \infty , \Delta z_{i} \rightarrow 0
    } \| \sum_{i=1}^{N} f( z_{i} ) \Delta z_{i} \| \\
     \leq \; \lim_{N \rightarrow \infty , \Delta z_{i} \rightarrow 0
    } \sum_{i=1}^{N} \| f( z_{i} ) \Delta z_{i} \| \; = \; \lim_{N \rightarrow \infty , \Delta z_{i} \rightarrow 0
    } \sum_{i=1}^{N}  \| f( z_{i} ) \| \| \Delta z_{i} \| \; \leq
    \; M \sum_{i=1}^{N} \| \Delta z_{i} \| \; = \; M L_{\gamma}
\end{multline}
\end{proof}

We are now ready to analyze a great difference among Complex
Calculus and Hyperbolic Calculus: in the latter case the Cauchy
Integral formula doesn't hold.

Let us recall that:
\begin{theorem} \label{th:complex Cauchy integral formula}
\end{theorem}
COMPLEX CAUCHY INTEGRAL FORMULA:

\begin{hypothesis}
\end{hypothesis}
\begin{equation}
    G  \subset {\mathbb{C}} \; : \; \pi_{1}(G) = \{ {\mathbb{I}} \}
    \; : \; \text{ f is analytic in G}
\end{equation}
\begin{equation}
    C \subset G \text{ closed smoooth curve not self-linking}
\end{equation}
\begin{equation}
    z_{0} \in Interior(\partial^{-1} C)
\end{equation}
\begin{thesis}
\end{thesis}
\begin{equation}
    f( z_{0} ) \; = \; \frac{1}{ 2 \pi j } \oint_{C} \frac{ f(z) }{ z - z_{0} }
\end{equation}
That theorem\ref{th:complex Cauchy integral formula} doesn't hold
if one replace $ {\mathbb{C}} $ with  $ {\mathbb{G}} $ may be
verified through the following:
\begin{example}
\end{example}
Let us consider the following  function  analytic in the whole
hyperbolic plane $f : {\mathbb{G}} \mapsto {\mathbb{G}}$:
\begin{equation}
    f(z) \; := \; z^{2} + a \; \; a \in {\mathbb{G}} : a \neq 0
\end{equation}
and the circle:
\begin{equation}
    c : [ 0 , 2 \pi ) \mapsto  {\mathbb{G}} \; : \; c(t) :=
    cos(t)+ i sin(t)
\end{equation}
One has that:
\begin{equation}
    f(0) \; = \; a \; \neq \; \frac{1}{2 \pi i}  \oint_{c} \frac{f(z)
    }{z-0} \; = \; \frac{1}{2 \pi i} ( \oint_{c} z + a \oint_{c}
    \frac{dz}{z}) \; = \;  \frac{a}{2 \pi i} ( \oint_{c} \frac{x}{x^{2} - y^{2}} dx - \frac{y}{x^{2} - y^{2}
    } dy \, + \, i \oint_{c} \frac{- y}{x^{2} - y^{2}} dx + \frac{x}{x^{2} -
    y^{2}} dy )
\end{equation}
\begin{equation}
    f(0) \; = \; a \; \neq \; \frac{a}{2 \pi i} \{    - 2 \int_{0}^{2 \pi} \frac{\sin(t) \cos(t)}{\cos^{2}(t) - \sin^{2} (t)
    } dt \, + i \, \int_{0}^{2 \pi} \frac{dt}{\cos^{2}(t) - \sin^{2}
    (t)} \} \; = \; - 2 I_{1} + i I_{2}
\end{equation}
where:
\begin{multline}
I_{1} \; := \; \int_{0}^{2 \pi} \frac{\sin(t) \cos(t)}{\cos^{2}(t)
- \sin^{2} (t)
    } dt \; = \; - \frac{1}{4} \lim_{\epsilon \rightarrow 0} ( [
\log | \cos (2 t )]_{0}^{\frac{\pi}{4}-\epsilon} + [ \log | \cos
(2 t )]_{\frac{\pi}{4}+\epsilon}^{\frac{3 \pi}{4}-\epsilon} + \\
[ \log | \cos (2 t )]_{\frac{3 \pi}{4}+\epsilon}^{\frac{5
\pi}{4}-\epsilon} + [ \log | \cos (2 t )]_{\frac{5
\pi}{4}+\epsilon}^{\frac{7 \pi}{4}-\epsilon} + [ \log | \cos (2 t
)]_{\frac{7 \pi}{4}+\epsilon}^{2 \pi} ) \;  = \; 0
\end{multline}
and:
\begin{multline}
    I_{2} \; := \; \int_{0}^{2 \pi} \frac{dt}{\cos^{2}(t) -
    \sin^{2}(t)} \; = \; \lim_{ \epsilon \rightarrow 0 } ([ arctanh
    ( \tan(t))]_{0}^{\frac{\pi}{4}-\epsilon} + [ arctanh
    ( \tan(t))]_{\frac{\pi}{4}+\epsilon}^{\frac{3 \pi}{4}-\epsilon}
    +  \\
    [ arctanh
    ( \tan(t))]_{\frac{3 \pi}{4}+\epsilon}^{\frac{5
    \pi}{4}-\epsilon}+[ arctanh
    ( \tan(t))]_{\frac{5 \pi}{4}+\epsilon}^{\frac{7
    \pi}{4}-\epsilon}+[ arctanh
    ( \tan(t))]_{\frac{7 \pi}{4}+\epsilon}^{2 \pi}) \; = \; 0
\end{multline}
so that:
\begin{equation}
    f(0) \; = \; a \; \neq \; 0
\end{equation}

\smallskip

 To understand the difference existing among the complex and
the hyperbolic cases let us recall the proof of
theorem\ref{th:complex Cauchy integral formula}:  in that one
consider the curve $ C' $ conciding with C apart from a little
deformation in which C go nearest to $ z_{0} $ along an horizontal
segment $ L_{1}$ , makes a circle $ \gamma_{\delta } $ of radius $
\delta $ with orientation opposite to that of C (clockwise) and
then returns to C along an horizontal segment $ L_{2} $ equal to $
L_{1} $ but with opposite direction.

In the complex case  one has that the function  $ \phi( z) \, :=
\, \frac{ f(z) - f ( z_{0}) }{ z - z_{0} } $ is analytic in the
region $ \partial^{-1} C' $ so that, by the Cauchy-Goursat
theorem, one has that:
\begin{equation}
 \oint_{C'} \phi( z) \; = \;  \oint_{C} \phi( z) \, + \, \oint_{L_{1}} \phi(
 z) \, + \, \int_{L_{2}} \phi(
 z) \, - \,  \oint_{\gamma_{\delta }} \phi( z) \; = \;  0
\end{equation}
and since:
\begin{equation}
 \int_{L_{2}} \phi(
 z) \; = \; - \int_{L_{1}} \phi(z)
\end{equation}
one can infer that:
\begin{equation}
      \oint_{C} \phi( z) \; = \; \oint_{\gamma_{\delta }} \phi( z)
\end{equation}
that applying the complex analogous of theorem\ref{lem:upper bound
to the integral along a finite curve} to the function $ \frac{f(z)
- f( z_{0})}{z - z_{0}}$ immediately leads to the thesis.

 Contrary, in the hyperbolic case, the function $ \phi( z)
\, := \, \frac{ f(z) }{ z - z_{0} } $ is not analytic in the
region $
\partial^{-1} C' $ since:
\begin{equation}
 \partial^{-1} C' \bigcap \{ x + i y \in {\mathbb{G}} \, : \, y -
 y_{0}  = \pm ( x - x_{0} ) \} \; \neq \; 0
\end{equation}

\begin{remark}
\end{remark}
One would be tempted to state the following:
\begin{conjecture} \label{con:first Khrennikov's conjecture}
\end{conjecture}
\begin{equation*}
    \int_{z z^{\star}=1} f (z) dz \; = \; 0 \; \; \forall f \;
    analytical
\end{equation*}
Conjecture\ref{con:first Khrennikov's conjecture} is, anyway,
false as it is shown by the following:
\begin{example}
\end{example}
Let us consider the following function $ f : {\mathbb{G}} \mapsto
{\mathbb{G}} $:
\begin{equation}
    f(z) \; := \; \exp(z) \; = \; u(x,y) \, + i \, v(x,y)
\end{equation}
\begin{equation}
    u(x,y) \; = \; \exp(x) \cosh(y)
\end{equation}
\begin{equation}
    v(x,y) \; = \; \exp(x) \sinh(y)
\end{equation}
and the two curves $ c_{i} : ( - \infty , + \infty ) \mapsto
{\mathbb{G}} \; i=1,2 $:
\begin{equation}
    c_{1} (t) \; = \; ( \cosh(t) , \sinh(t) )
\end{equation}
\begin{equation}
    c_{2} (t) \; = \; ( - \cosh(t) , - \sinh(t) )
\end{equation}
One has that:
\begin{equation}
    \int_{c_{1}} f(z) dz \; = \; \int_{c_{1}} \exp(x) \cosh(y) dx
    + \exp(x) \sinh (y) dy \, + \, i \int_{c_{1}} \exp(x) \sinh(y)
    dx + \exp(x) \cosh(y) dy
\end{equation}
and hence:
\begin{multline}
 \int_{c_{1}} f(z) dz \; = \; \int_{- \infty}^{+ \infty} \exp (
 \cosh(t)) \cosh( \sinh(t)) + \exp (
 \cosh(t)) \sinh ( \sinh(t)) \cosh(t) \\
  + \, i \int_{- \infty}^{+
 \infty} \exp (
 \cosh(t)) \sinh ( \sinh(t)) \sinh(t) + \exp (
 \cosh(t)) \cosh( \sinh (t)) \cosh(t) dt \\ = \; [ \exp ( \cosh
 (t))  \cosh ( \sinh(t))]_{- \infty}^{+ \infty} + i [
 \exp( \cosh (t) ) \sinh( \sinh(t)) ]_{-\infty}^{+ \infty}
\end{multline}
and:
\begin{equation}
    \int_{c_{2}} f(z) dz \; = \; \int_{c_{2}} \exp(x) \cosh(y) dx
    + \exp(x) \sinh (y) dy \, + \, i \int_{c_{2}} \exp(x) \sinh(y)
    dx + \exp(x) \cosh(y) dy
\end{equation}
and hence:
\begin{multline}
 \int_{c_{2}} f(z) dz \; = \; \int_{- \infty}^{+ \infty} \exp (
 - \cosh(t)) \cosh(- \sinh(t)) + \exp (
 - \cosh(t)) \sinh ( - \sinh(t)) ( - \cosh(t) ) \\
  + \, i \int_{- \infty}^{+
 \infty} \exp (
- \cosh(t)) \sinh (- \sinh(t)) (- \sinh(t)) + \exp (
 - \cosh(t)) \cosh( - \sinh (t)) ( - \cosh(t) ) dt \\ = \;
  [ \exp (- \cosh
 (t))  \cosh ( \sinh(t))]_{- \infty}^{+ \infty} + i [-
 \exp( - \cosh (t) ) \sinh( \sinh(t)) ]_{-\infty}^{+ \infty}
\end{multline}
from which it follows that:
\begin{equation}
    \int_{c_{1}} f(z) dz \; \pm \;  \int_{c_{2}} f(z) dz \; \neq
    \; 0
\end{equation}

\smallskip

\begin{example}
\end{example}
Let us consider the following function $ f : {\mathbb{G}} \mapsto
{\mathbb{G}} $:
\begin{equation}
    f(z) \; := \; z^{2} \; = \; u(x,y) \, + i \, v(x,y)
\end{equation}
\begin{equation}
    u(x,y) \; = \; x^{2} + y^{2}
\end{equation}
\begin{equation}
    v(x,y) \; = \; 2 x y
\end{equation}
and the two curves $ c_{i} : ( - \infty , + \infty ) \mapsto
{\mathbb{G}} \; i=1,2 $:
\begin{equation}
    c_{1} (t) \; = \; ( \cosh(t) , \sinh(t) )
\end{equation}
\begin{equation}
    c_{2} (t) \; = \; ( - \cosh(t) , - \sinh(t) )
\end{equation}
One has that:
\begin{equation}
    \int_{c_{1}} z^{2} dz \; = \; \int_{c_{1}} ( x^{2} + y^{2} )
    dx + 2 x y dy \, + \, i \int_{c_{1}}  2 x y dx + ( x^{2} + y^{2} )
    dy
\end{equation}
and hence:
\begin{multline}
  \int_{c_{1}} z^{2} dz \; = \; \int_{- \infty}^{+ \infty} (
  \cosh^{2}(t) + \sinh^{2} (t) ) \sinh (t) + 2 \cosh^{2}(t)
  \sinh(t) dt \\
  \, + i \int_{- \infty}^{+ \infty} 2 \cosh(t) \sinh^{2}(t) + (
  \cosh^{2}(t) + \sinh^{2} (t) ) \cosh(t) d t \\
  = \; [ \frac{1}{3} \cosh (t) ]_{- \infty}^{+ \infty} + i  [ \frac{1}{3} \sinh (t) ]_{- \infty}^{+ \infty}
\end{multline}
and:
\begin{equation}
    \int_{c_{2}} z^{2} dz \; = \; \int_{c_{2}} ( x^{2} + y^{2} )
    dx + 2 x y dy \, + \, i \int_{c_{1}}  2 x y dx + ( x^{2} + y^{2} )
    dy
\end{equation}
and hence:
\begin{multline}
  \int_{c_{2}} z^{2} dz \; = \; \int_{- \infty}^{+ \infty} (
  \cosh^{2}(t) + \sinh^{2} (t) ) (- \sinh (t)) + 2 \cosh^{2}(t)
  ( - \sinh(t)) dt \\
  \, + i \int_{- \infty}^{+ \infty} 2 \cosh(t) ( - \sinh^{2}(t) ) + (
  \cosh^{2}(t) + \sinh^{2} (t) ) ( - \cosh(t) ) d t \\
  = \; [ - \frac{1}{3} \cosh (3 t) ]_{- \infty}^{+ \infty} + i  [ - \frac{1}{3} \sinh (t) ]_{- \infty}^{+ \infty}
\end{multline}from which it follows that:
\begin{equation}
    \int_{c_{1}} f(z) dz \; \pm \;  \int_{c_{2}} f(z) dz \; \neq
    \; 0
\end{equation}

\smallskip

With analogy to Conjecture\ref{con:first Khrennikov's conjecture}
one would be tempted to state the following:
\begin{conjecture} \label{con:second Khrennikov's conjecture}
\end{conjecture}
\begin{equation*}
    f( z_{0} ) \; = \; \frac{1}{2 \pi i} \int_{ (z-z_{0}) ( z^{\star} - z_{0}^{\star} ) = 1
    } \frac{ f(z) }{ z - z_{0} } \; \; \forall f \text{ analytic}
\end{equation*}
Conjecture\ref{con:second Khrennikov's conjecture} is, anyway,
false as it is proved by the following:
\begin{example}
\end{example}
Given the function:
\begin{equation}
    f(z) \; := \; z^{2} + c \; \; c \neq 0
\end{equation}
and the two curves $ c_{i} : ( - \infty , + \infty ) \mapsto
{\mathbb{G}} \; i=1,2 $:
\begin{equation}
    c_{1} (t) \; = \; ( \cosh(t) , \sinh(t) )
\end{equation}
\begin{equation}
    c_{2} (t) \; = \; ( - \cosh(t) , - \sinh(t) )
\end{equation}
we will show that:
\begin{equation}
    \frac{1}{2 \pi i} ( \int_{c_{1}} \frac{ f(z)}{z} \pm  \int_{c_{2}} \frac{
    f(z)}{z} ) \; \neq \; 0
\end{equation}
At this purpose let us observe that:
\begin{multline}
 \frac{1}{2 \pi i} ( \int_{c_{1}} \frac{z^{2}+c}{z} \pm \int_{c_{2}}
 \frac{z^{2}+c}{z}) \; = \\
  \frac{1}{2 \pi i} (  \int_{c_{1}} x dx + y d y   + i
 \int_{c_{1}} y dx + x dy + c ( \int_{c_{1}} \frac{x}{x^{2}-y^{2}}
 dx - \frac{y}{x^{2}-y^{2}} d y + i \int_{c_{1}} \frac{-
 y}{x^{2}-y^{2}} dx + \frac{x}{x^{2}-y^{2}} dy )) \\
  \pm \frac{1}{2 \pi i} (  \int_{c_{2}} x dx + y d y   + i
 \int_{c_{2}} y dx + x dy + c ( \int_{c_{2}} \frac{x}{x^{2}-y^{2}}
 dx - \frac{y}{x^{2}-y^{2}} d y + i \int_{c_{2}} \frac{-
 y}{x^{2}-y^{2}} dx + \frac{x}{x^{2}-y^{2}} dy ))
\end{multline}
and hence:
\begin{multline}
 \frac{1}{2 \pi i} ( \int_{c_{1}} \frac{z^{2}+c}{z} \pm \int_{c_{2}}
 \frac{z^{2}+c}{z}) \; = \\
\frac{1}{2 \pi i} ( \int_{- \infty}^{+ \infty} 2 \cosh(t) \sinh(t)
\pm  \int_{- \infty}^{+ \infty} 2 \cosh(t) \sinh(t) +  i ( \int_{-
\infty}^{+ \infty} dt ( \sinh^{2} (t) + \cosh^{2} (t)) \\
\pm \int_{- \infty}^{+ \infty} dt ( \sinh^{2} (t) + \cosh^{2}
(t)))) + \frac{c}{2 \pi } ( \int_{- \infty}^{+ \infty} dt \pm
\int_{- \infty}^{+ \infty} dt )
\end{multline}
so that:
\begin{equation}
    \frac{1}{2 \pi i} ( \int_{c_{1}} \frac{z^{2}+c}{z} + \int_{c_{2}}
 \frac{z^{2}+c}{z}) \; = + \infty \; \neq \; c
\end{equation}
\begin{equation}
    \frac{1}{2 \pi i} ( \int_{c_{1}} \frac{z^{2}+c}{z} - \int_{c_{2}}
 \frac{z^{2}+c}{z}) \; = \; 0 \; \neq \; c
\end{equation}
\newpage
\section{Multivalued functions on the hyperbolic plane and hyperbolic riemann surfaces}

The introduction of multi-valued functions in Complex Calculus
realizes an unexpected bridge between Analysis and Differential
Geometry through the double nature of the notion of riemann
surface as a collection  of riemann sheets \cite{Hassani-99} and
as a one-dimensional compact orientable complex manifold
\cite{Nakahara-95}.

Following \cite{Sobczyk-95} let us start introducing the following
map:
\begin{definition}
\end{definition}
$|\| \cdot |\| : {\mathbb{G}} \rightarrow [0,\infty )$:
\begin{equation}
 |\| x + i y |\| \; := \; \sqrt{| x^{2} - y^{2} | }
\end{equation}

drawn in figure\ref{fig:Sobczyk modulus}:
\begin{figure} [h]
  \includegraphics[width=150pt]{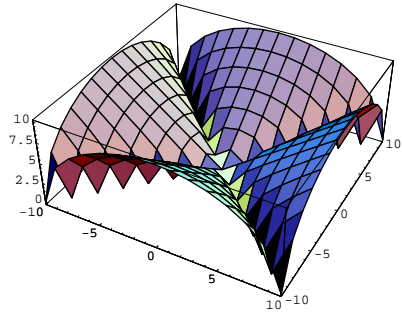}\\
  \caption{the map $  |\| \cdot  |\|  $}\label{fig:Sobczyk modulus}
\end{figure}

Clearly:
\begin{proposition}
\end{proposition}
\begin{equation*}
   |\| z |\| \; = \; | z | \; \; \forall z \in {\mathbb{G}}_{+}
\end{equation*}

Given $ r \in [0,\infty ) $:
\begin{definition}
\end{definition}
\begin{equation*}
  Hyp_{r} \; := \; \{ z \in {\mathbb{G}} \; : \; |\| z |\| \, = \,
  r \}
\end{equation*}
Clearly:
\begin{theorem}
\end{theorem}
\begin{equation*}
  Hyp_{0} \; = \; Diagonals
\end{equation*}

The set $ Hyp_{1}$ is drawn in figure\ref{fig:hyperbolas}:

\begin{figure} [h]
  \includegraphics{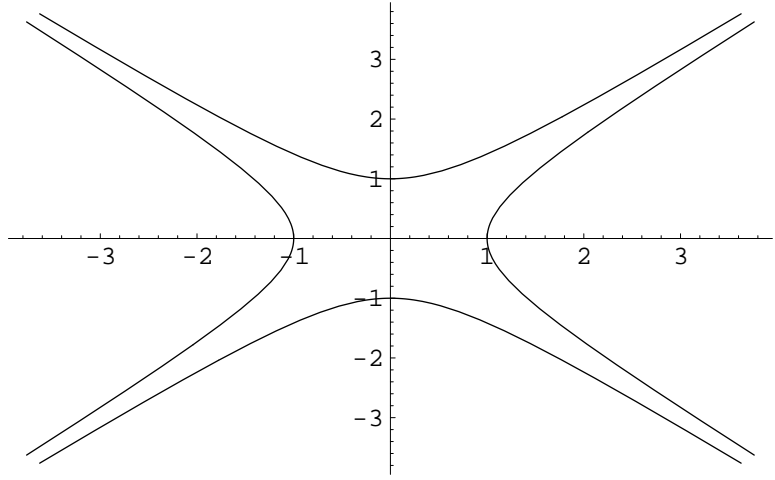}\\
  \caption{The set $ Hyp_{1}$} \label{fig:hyperbolas}
\end{figure}

One has clearly that:
\begin{theorem}
\end{theorem}
\begin{equation*}
 {\mathbb{G}} \; = \; \cup_{r \in  [0,\infty )} Hyp_{r}
\end{equation*}
Furthermore introduced the following:
\begin{definition}
\end{definition}
HYPERBOLIC QUADRANTS
\begin{equation*}
    H_{1} \; := \; \{ x+iy \in {\mathbb{G}} \, : \, | y | < x , x
    > 0 \}
\end{equation*}
\begin{equation*}
    H_{2} \; := \; \{ x+iy \in {\mathbb{G}} \, : \, | y | > x , y
    > 0 \}
\end{equation*}
\begin{equation*}
    H_{3} \; := \; \{ x+iy \in {\mathbb{G}} \, : \, | y | < x , x
    < 0 \}
\end{equation*}
\begin{equation*}
    H_{4} \; := \; \{ x+iy \in {\mathbb{G}} \, : \, | y | > x , y
    < 0 \}
\end{equation*}
one has clearly that:
\begin{theorem}
\end{theorem}
\begin{equation*}
    {\mathbb{G}} \; = \; \bigcup_{i=1}^{4} H_{i} \cup Diagonals
\end{equation*}
\begin{equation*}
     {\mathbb{G}}_{+}^{\star} \; = \; H_{1} \cup H_{3}
\end{equation*}
Introduced the following:
\begin{definition} \label{def:hyperbolic argument}
\end{definition}
HYPERBOLIC ARGUMENT OF $z= x+ iy \in {\mathbb{G}} - Diagonals $:
\begin{equation*}
    \theta( z) \; := \; \left\{%
\begin{array}{ll}
    arctanh ( \frac{y}{x} ), & \hbox{if $z \in H_{1} \cup H_{3}$ ;} \\
    arctanh ( \frac{x}{y} ), & \hbox{if $z \in H_{2} \cup H_{4}$ .} \\
\end{array}%
\right.
\end{equation*}
we can finally state the following:
\begin{theorem}
\end{theorem}
EXPONENTIAL REPRESENTATION OF $z= x+ iy \in {\mathbb{G}} -
Diagonals $:
\begin{equation*}
    z \; = \; \left\{%
\begin{array}{ll}
    r \exp ( i \theta ) , & \hbox{if $ z \in H_{1}$;} \\
     i r \exp ( i \theta ) , & \hbox{if $ z \in H_{2}$;} \\
    - r \exp ( i \theta ) , & \hbox{if $ z \in H_{3}$;} \\
     - i r \exp ( i \theta ) , & \hbox{if $ z \in H_{4}$ .} \\
\end{array}%
\right.
\end{equation*}
Let us recall, now, that in Complex Analysis,  given a map $ f :
{\mathbb{C}} \mapsto {\mathbb{C}} $ and a point $ z_{0} \in
{\mathbb{C}} $:
\begin{definition} \label{def:branch point in Complex Analysis}
\end{definition}
 $ z_{0} $ IS A BRANCH POINT OF f:
\begin{equation*}
    f( r_{0} , \theta_{0} ) \; \neq \;   f( r_{0} , \theta_{0} + 2 \pi
    ) \text{ for every closed curve C encircling  $ z_{0} $}
\end{equation*}
\begin{remark}
\end{remark}
 Branch points emerge in Complex Calculus owing to the
arbitrariness, up to a multiple of $ [ 0 , 2 \pi) $ of the
argument.

Since such an arbitrariness doesn't occur as to the hyperbolic
argument, the definition\ref{def:branch point in Complex Analysis}
has no analogue in Hyperbolic Analysis where the source of
multi-valued functions  lies elsewhere.

\begin{example}
\end{example}
MULTIVOCITY OF THE SQUARE-ROOT AND ITS CONSEQUENCES

Given $ z_{1} , z_{2} \in {\mathbb{G}}  $:
\begin{definition}
\end{definition}
$ z_{1} $ IS A SQUARE-ROOT OF $ z_{2} \; ( z_{1} \, = \, \sqrt{
z_{2} }) $
\begin{equation}
 z_{1}^{2} \; = \; z_{2}
\end{equation}
One has clearly that:
\begin{lemma} \label{le:on the bivocity of the square root}
\end{lemma}
ON THE BIVOCITY OF THE SQUARE ROOT:
\begin{equation}
    z_{1} \, = \, \sqrt{
z_{2} } \; \Leftrightarrow \;  i \, z_{1} \, = \, \sqrt{ z_{2} }
\end{equation}
\begin{lemma} \label{le:on the non-existence of the square-root of a negative number}
\end{lemma}
ON THE NON-EXISTENCE OF THE SQUARE ROOT OF A NEGATIVE NUMBER
\begin{equation}
    \nexists \sqrt{ - x} \; \; \forall x \in {\mathbb{R}}_{+}
\end{equation}

\smallskip

 Lemma\ref{le:on the bivocity of the square root} and
Lemma\ref{le:on the non-existence of the square-root of a negative
number} imply the following
\begin{theorem}
\end{theorem}
NO-GO FUNDAMENTAL THEOREM OF HYPERBOLIC ALGEBRA

\begin{hypothesis}
\end{hypothesis}

\begin{equation}
    a_{0} , \cdots , a_{n} \in {\mathbb{G}}
\end{equation}

\begin{thesis}
\end{thesis}
\begin{equation}
 \neg ( \exists ! ( z_{1} \, \cdots \, z_{n} ) \in
{\mathbb{G}}^{n} \, : \, \sum_{i=0}^{n} a_{i} z^{i} \, = \, 0 )
\end{equation}

\begin{example}
\end{example}
SECOND DEGREE EQUATIONS ON THE HYPERBOLIC PLANE

Given the equation:
\begin{equation} \label{eq:second degree equation}
    a z^{2} + b z + c = 0 \; \; a ,b,c \in {\mathbb{G}} \, a \neq
    0
\end{equation}

  one has that:
\begin{enumerate}
    \item if the discriminant $ \Delta := b^{2} - 4  a c \;> \; 0
    $ then eq.\ref{eq:second degree equation} has 4 solutions:
    \begin{eqnarray}
      z_{1 } \;& =& \; \frac{ - b + (\sqrt{\Delta})_{1} }{ 2 a } \\
      z_{2 } \;& =& \; \frac{ - b - (\sqrt{\Delta})_{1} }{ 2 a } \\
      z_{3 } \;& =& \; \frac{ - b + (\sqrt{\Delta})_{2} }{ 2 a } \\
      z_{4 } \;& =& \; \frac{ - b - (\sqrt{\Delta})_{2} }{ 2 a } \\
    \end{eqnarray}
    where $ (\sqrt{\cdot}) _{i} $ denote the $ i^{th}$ branch of the
    square root $ i=1,2 $.
    \item if the discriminant $ \Delta := b^{2} - 4  a c \; = \;
    0 $ then eq.\ref{eq:second degree equation} has 1 solutions:
\begin{equation}
    z_{1} \; = \; \frac{-b}{2 a}
\end{equation}
    \item  if the discriminant $ \Delta := b^{2} - 4  a c \; < \; 0
    $ then eq.\ref{eq:second degree equation} has no solution
\end{enumerate}

\smallskip

As in the complex case it appears natural to introduce
(hyperbolic) Riemannian surfaces for multi-valued functions over $
{\mathbb{G}}$.

\smallskip

\begin{example}
\end{example}
THE HYPERBOLIC RIEMANN SURFACE OF THE SQUARE ROOT:

the hyperbolic riemannian surface associated to the square roots:
it is made of four sheets $ \{ {\mathbb{G}}^{(i)} \}_{i=1}^{4} $
such that:
\begin{enumerate}
    \item  any $ {\mathbb{G}}^{(i)} $  is
cutted along the semi-axis $  {\mathbb{R}}_{+} $
    \item the lower border of the cut in $
{\mathbb{G}}^{(1)} $ is pasted with the upper border of the cut in
$ {\mathbb{G}}^{(2)}$
    \item the upper border of the cut in $
{\mathbb{G}}^{(2)} $ is pasted with the lower border of the cut in
$ {\mathbb{G}}^{(3)}$
  \item the upper border of the cut in $
{\mathbb{G}}^{(3)} $ is pasted with the lower border of the cut in
$ {\mathbb{G}}^{(4)}$
 \item the upper border of the cut in $
{\mathbb{G}}^{(4)} $ is pasted with the lower border of the cut in
$ {\mathbb{G}}^{(1)}$
\begin{equation}
 \sqrt{x} \in {\mathbb{R}}^{(1)}_{+} \; \; \forall x \in  {\mathbb{R}}^{(1)}_{+}
\end{equation}
 \item
\begin{equation}
 \sqrt{x} \in  {\mathbb{R}}^{(2)}_{-} \; \; \forall x \in  {\mathbb{R}}^{(2)}_{+}
\end{equation}
 \item
\begin{equation}
 \sqrt{x} \in  i{\mathbb{R}}^{(3)}_{+} \; \; \forall x \in  {\mathbb{R}}^{(3)}_{+}
\end{equation}
 \item
\begin{equation}
 \sqrt{x} \in  i{\mathbb{R}}^{(4)}_{-} \; \; \forall x \in  {\mathbb{R}}^{(4)}_{+}
\end{equation}
  \item if z starts from $
z_{0} \in {\mathbb{G}}^{(i)} $ and describes a closed contour
containing the origin , then $ \sqrt{z} $ passes from the $ i^{th}
$ sheet  to the $ (i+1)^{th}$ sheet and thus the point on the
hyperbolic Riemann surface passes from $ {\mathbb{G}}^{(i)} $ to $
{\mathbb{G}}^{(i+1)} $ for i=1,2,3.
 \item if z starts from $
z_{0} \in {\mathbb{G}}^{(4)} $ and describes a closed contour
containing the origin , then $ \sqrt{z} $ passes from the $ 4^{th}
$ sheet  to the $ 1^{th}$ sheet and thus the point on the
hyperbolic Riemann surface passes from $ {\mathbb{G}}^{(4)} $ to $
{\mathbb{G}}^{(1)} $.
\end{enumerate}

\smallskip

Let us now look at hyperbolic riemann surfaces from a
differential-geometric viewpoint.

 Given a topological space M:
\begin{definition}
\end{definition}
HYPERBOLIC ANALYTIC ATLAS ON M

a family $ \{ ( U_{\alpha} , \phi_{\alpha} ) \}_{\alpha \in I} $
(where I is some arbitrary index set) such that:
\begin{enumerate}
    \item $ \{ U_{\alpha} \}_{\alpha \in I} $  is an open covering
    of M, i.e. $ \cup_{\alpha \in I} U_{\alpha} = M$
    \item $ \phi_{\alpha} $ is an homeomorphism from $  U_{\alpha}
    $ to an open subset $ U_{\alpha}' $ of $ {\mathbb{G}}^{m} $,
    \item Given $ U_{\alpha} , U_{\beta} $ such that :  $ U_{\alpha} \bigcap
    U_{\beta} \neq \emptyset $ the map $ \psi_{\beta \alpha} : \phi_{\alpha} ( U_{\alpha} \cap U_{\beta} ) \mapsto
      \phi_{\beta} ( U_{\alpha} \cap U_{\beta} )$ :
\begin{equation*}
   \psi_{\beta \alpha}  \; := \; \phi_{\beta} \phi_{\alpha}^{- 1}
\end{equation*}
is analytic.
\end{enumerate}
We will denote the set of all the hyperbolic analytic atlas of M
as HYP-ATLAS(M).

Given $ x_{1} , x_{2} \in HYP-ATLAS(M) $:
\begin{definition}
\end{definition}
$ x_{1} $ AND $ x_{2} $ ARE COMPATIBLE ( $ x_{1} \sim_{C} x_{2}
$):
\begin{equation*}
    x_{1} \cup x_{2} \in HYP-ATLAS(M)
\end{equation*}
It may be easily proved that $ \sim_{C} $ is an equivalence
relation.
\begin{definition}
\end{definition}
HYPERBOLIC STRUCTURES OVER M:
\begin{equation}
    HS ( M) \; := \; \frac{HYP-ATLAS(M)}{\sim_{C}}
\end{equation}
M endowed with an hyperbolic structure will be called an
hyperbolic manifold.

\begin{definition}
\end{definition}
HYPERBOLIC RIEMANN SURFACE

a one-dimensional compact orientable hyperbolic manifold

\smallskip

\begin{example}
\end{example}
HYPERBOLIC STRUCTURES ON THE TORUS:

Let us consider two hyperbolic numbers $ \omega_{1} , \omega_{2}
\in {\mathbb{G}} $ such  that:
\begin{equation}
    \frac{\omega_{2}}{\omega_{1}} \notin {\mathbb{R}} \; and \; Im
    ( \frac{\omega_{2}}{\omega_{1}}) > 0
\end{equation}
Introduced the lattice:
\begin{equation}
    L( \omega_{1} , \omega_{2} ) \; = \{ n_{1} \omega_{1} +  n_{2}
    \omega_{2} \; \; n_{1}, n_{2} \in {\mathbb{Z}} \}
\end{equation}
let us observe that the hyperbolic structure of $ {\mathbb{G}} $
induces an hyperbolic structure over $ \frac{{\mathbb{G}}}{L} \; =
\; T^{(2)} $.

Let us observe that there are many pairs $ ( \omega_{1},
\omega_{2}) $ which give rise to the same hyperbolic structure  on
$  T^{(2)} $.

The problem of the classification of the hyperbolic structures of
the torus is under investigation.

\smallskip

\begin{example}
\end{example}
Let us consider the stereographic coordinates of a point $
P(x,y,z) \in S^{2} - \{ North \; Pole \} $ projected from the
North Pole:
\begin{equation}
    (X,Y) \; := \; (  \frac{x}{1-z} , \frac{y}{1-z} )
\end{equation}
and the stereographic coordinates of a point $ P(x,y,z) \in  S^{2}
- \{ South \; Pole \}$ projected from the South Pole:
\begin{equation}
    ( U, V) \; := \; ( \frac{x}{1+z} , \frac{- y }{ 1+z} )
\end{equation}
Introduced the hyperbolic coordinates:
\begin{eqnarray}
  Z \; :&=& \; X + i Y \\
  W \; :&=& \; U - i V
\end{eqnarray}
one has that:
\begin{equation}
    W = \frac{x - i y }{1+z} \; = \; \frac{1-z}{1+z} ( X - i Y) \; = \;
    \frac{X-iY}{X^{2}+Y^{2}} \; \neq \; \frac{1}{Z} \; = \; \frac{X-iY}{X^{2} - Y^{2}}
\end{equation}
Let us now observe that the function $ F(Z) = \alpha (X,Y) + i
\beta (X,Y) $:
\begin{equation}
     \alpha (X,Y) \; := \; \frac{X}{X^{2}+Y^{2}}
\end{equation}
\begin{equation}
  \beta (X,Y) \; := \;\frac{-Y}{X^{2}+Y^{2}}
\end{equation}
doesn't obey the hyperbolic Cauchy-Riemann condition and isn't
hence an hyperbolic analytic function.

So stereographic coordinates don't allow to define an hyperbolic
structure on $ S^{(2)} $.

Let us consider instead the hyperboloid:
\begin{equation}
    H^{(2)} \; := \; \{ ( x,y,z) \in {\mathbb{R}}^{3} \, : \,
    x^{2}-y^{2}+z^{2} \, = \, 1 \}
\end{equation}
Introduced again the coordinates:
\begin{equation}
    (X,Y) \; := \; (  \frac{x}{1-z} , \frac{y}{1-z} )
\end{equation}
\begin{equation}
    ( U, V) \; := \; ( \frac{x}{1+z} , \frac{- y }{ 1+z} )
\end{equation}
\begin{eqnarray}
  Z \; :&=& \; X + i Y \\
  W \; :&=& \; U - i V
\end{eqnarray}
one has that:
\begin{equation}
    W = \frac{x - i y }{1+z} \; = \; \frac{1-z}{1+z} ( X - i Y) \; = \;
    \frac{X-iY}{X^{2}-Y^{2}} \; = \; \frac{1}{Z}
\end{equation}
So we have that W is an analytical function everywhere in $
{\mathbb{R}}^{3} $ outside $ \Delta := \{ (x,y,z) \in
{\mathbb{R}}^{3} \, : \, y = \pm x \} $

Let us introduce the set:
\begin{equation}
   H^{(2)}_{cut} \; := \; H^{(2)} -  \Delta
\end{equation}
In \cite{Motter-Rosa-98} A.E. Motter and M.A.F. Rosa endow $
H^{(2)} $ with an hyperbolic structure and propose the resulting
hyperbolic manifold  as the natural candidate to the role of
Hyperbolic Riemann Sphere defying the readers to make a better
proposal.

According to our modest opinion what is lacking in Motter's and
Rosa's proposal is a clear illustration on how the infinity point
emerges from their stuff.

This is not the case as to the hyperbolic manifold $ H^{(2)}_{cut}
$ endowed with the hyperbolic structure induced by the previously
defined coordinates Z and W.

Since such an hyperbolic manifold may be identified with  $
{\mathbb{G}} \cup \{ \infty \} - \Delta $ we propose it as
alternative candidate to the role of Hyperbolic Riemann Sphere.

\newpage
\section{Physical application to the vibrating string}
Let $ f : S \mapsto {\mathbb{G}} $ be a function analytic in its
domain of definition $ S \subset {\mathbb{G}} $ that we will
suppose to be connected and simply-connected.

By the Corollary\ref{cor:real and imaginary part obey the wave
equation} it follows that that the real and imaginary parts of $ f
( z) \, = \; f_{1}(x,y) + i f_{2}(x,y) $ obey on S   the
one-dimensional wave equation:
\begin{equation}
     ( \partial_{t}^{2} - \partial_{x}^{2} ) f_{i}(x,y) \; = \; 0
     \; \; i=1,2
\end{equation}
This implies that, on S $ f_{i}(x,y) \, i=1,2 $ is of the form:
\begin{equation}
  f_{i} ( x,y) \; = \; c_{i,1} g_{i}( x+y) + c_{i,2} g_{i}( x-y) \; \; c_{i,1}, c_{i,2} \in
  {\mathbb{R}} , i=1,2
\end{equation}
where $ g_{i } \in C^{1} ( {\mathbb{R}} ) \; i=1,2 $.

Furthermore the solution of the Cauchy problem for the
1-dimensional wave equation \cite{Rubinstein-Rubinstein-98}
implies that if we know the value and the rate of change of  $
f_{i}$  on one of the coordinates axes we can infer the values of
$ f_{i}$ on the whole hyperbolic plane in the following way:
\begin{equation}
    [( f_{i}(0,y) \, = \, g_{i}(y) ) \: \wedge \: (\partial_{y}
     f_{i}(0,y) \, = \,  h_{i}(y)) \; \Rightarrow \; ( f_{i}(x,y) \; = \; \frac{1}{2} ( g_{i}( y-x)+ g_{i}( y+x) ) +
    \int_{y-x}^{y+x} h_{i} (s) d s )]  \; \; \forall g_{i}, h_{i} \in C^{1} ( {\mathbb{R}}
    ) , i=1,2
\end{equation}
\begin{equation}
    [( f_{i}(x,0) \, = \, g_{i}(x)) \: \wedge \: ( \partial_{x}
     f_{i}(x,0) \, = \,  h_{i}(x)) \; \Rightarrow \; ( f_{i}(x,y) \; = \; \frac{1}{2} ( g_{i}( x-y)+ g_{i}( x+y) ) +
    \int_{x-y}^{x+y} h_{i} (s) d s )] \; \; \forall g_{i}, h_{i} \in C^{1} ( {\mathbb{R}}
    ) , i=1,2
\end{equation}

\smallskip

As in the complex case the fact that the real and imaginary part
of an analytic function obey the Laplace equation may be used to
apply Complex Analysis to Electrostatics, in the hyperbolic case
the fact that the real and imaginary part of an analytic function
obey the 1-dimensional wave equation should allow to apply
Hyperbolic Analysis to the physics of a vibrating string.

\newpage
\section{Hyperbolic Analysis as the (1,0)-case of Clifford Analysis}
It is interesting to investigate whether the formalization of
Hyperbolic Analysis performed in the previous sections is
compatible with the more general Clifford calculus.

Following \cite{Gurlebeck-Sprossig-97} let us introduce first of
all the following notation:
\begin{definition}
\end{definition}
\begin{equation}
    {\mathbb{R}}^{(p,q)} \; := \; ( {\mathbb{R}}^{p+q} , \tilde{q} ) \: :
    \: sign(\tilde{q}) = (p,q)
\end{equation}
Denoted with $ \{ \hat{e}_{1} , \cdots , \hat{e}_{p+q} \} $ the
canonical basis of $ {\mathbb{R}}^{(p,q)} $ one introduces the
following operators:
\begin{definition}
\end{definition}
DIRAC OPERATOR ON  $ {\mathbb{R}}^{(p,q)} $:
\begin{equation}
    D : C^{1}( {\mathbb{R} }^{(p,q)} , Cl_{p,q}  ) \, \mapsto \, C^{1}( {\mathbb{R} }^{(p,q)} , Cl_{p,q}
    ) \; : \; D := \sum_{i=1}^{p+q} \hat{e}_{i} \partial_{i}
\end{equation}
\begin{definition}
\end{definition}
CAUCHY-FUETER OPERATOR ON $  {\mathbb{R}} \otimes
{\mathbb{R}}^{(p,q)} $:
\begin{equation}
 \partial :  C^{1}({\mathbb{R}} \otimes {\mathbb{R} }^{(p,q)} , Cl_{p,q}  ) \, \mapsto \, C^{1}( {\mathbb{R}} \otimes {\mathbb{R} }^{(p,q)} , Cl_{p,q}
    ) \; : \; \partial := \partial_{0} + D
\end{equation}
\begin{definition}
\end{definition}
ADJOINT DIRAC OPERATOR ON  $ {\mathbb{R}}^{(p,q)} $:
\begin{equation}
    \bar{D} : C^{1}( {\mathbb{R} }^{(p,q)} , Cl_{p,q}  ) \, \mapsto \, C^{1}( {\mathbb{R} }^{(p,q)} , Cl_{p,q}
    ) \; : \; \bar{D} := \sum_{i=1}^{p+q} \bar{\hat{e}}_{i} \partial_{i}
\end{equation}
where:
\begin{equation}
    \bar{x} \; := \; (-1)^{\frac{ degree(x)( degree(x)+1)}{2}} \; \;
\end{equation}
is the conjugate of $ x \in Cl_{p,q}$.
\begin{definition}
\end{definition}
ADJOINT CAUCHY-FUETER OPERATOR ON $  {\mathbb{R}} \otimes
{\mathbb{R}}^{(p,q)} $:
\begin{equation}
\bar{ \partial} :  C^{1}({\mathbb{R}} \otimes {\mathbb{R}
}^{(p,q)} , Cl_{p,q}  ) \, \mapsto \, C^{1}( {\mathbb{R}} \otimes
{\mathbb{R} }^{(p,q)} , Cl_{p,q}
    ) \; : \; \bar{\partial} := \partial_{0} - D
\end{equation}
Given $ f  \in C^{1}( {\mathbb{R} }^{(p,q)} , Cl_{p,q}  ) $:
\begin{definition}
\end{definition}
f IS $ CL_{p,q}$-REGULAR:
\begin{equation}
    D f \; = \; 0
\end{equation}
Given $ f  \in C^{1}( {\mathbb{R}} \otimes {\mathbb{R} }^{(p,q)} ,
Cl_{p,q}  ) $:
\begin{definition}
\end{definition}
f IS $ CL_{p,q}$-HOLOMORPHIC
\begin{equation}
   \partial f \; = \; 0
\end{equation}

One has that \cite{Gurlebeck-Sprossig-97}:
\begin{theorem}
\end{theorem}
GURLEBECK-SPROSSIG'S (0,n)-CAUCHY GOURSAT THEOREM:

\begin{hypothesis}
\end{hypothesis}
\begin{equation*}
 G \subset {\mathbb{R}}^{n}
\end{equation*}
\begin{equation*}
    u \in C^{1} ( G , Cl_{0,n} ) \cap C( \bar{G},Cl_{0,n} )
\end{equation*}
\begin{equation*}
    u \in Ker D
\end{equation*}
\begin{equation*}
    S \subset G \text{ surface}
\end{equation*}
\begin{equation*}
    n_{S}(y) \text{ outward pointing unit normal to S at y }
\end{equation*}
\begin{thesis}
\end{thesis}
\begin{equation*}
\int_{S} u(y) n_{S}(y)  dS_{y} \; = \; 0
\end{equation*}

\smallskip

Let us now introduce the following maps:
\begin{definition}
\end{definition}
SURFACE AREA OF THE UNIT SPHERE IN $  {\mathbb{R}}^{n} $:
\begin{equation*}
    \sigma_{n} \; := \; \int_{S^{n-1}} d S  \; = \; \frac{ 2 \pi^{\frac{n}{2}}
    }{\Gamma(\frac{n}{2})}
\end{equation*}
where $ \Gamma ( x) \; := \; \int_{0}^{\infty} \exp (t) t^{x-1} dt
$ is Euler's Gamma function.

Assumed that $ n > 2 $:
\begin{definition}
\end{definition}
$ E : {\mathbb{R}}^{n} \rightarrow  {\mathbb{R}} $:
\begin{equation*}
    E(x) \; := \; \frac{1}{ \sigma_{n} }  \frac{1}{2 - n} | x |^{-(n-2)}
\end{equation*}
\begin{definition}
\end{definition}
$ e : {\mathbb{R}}^{n} \rightarrow  {\mathbb{R}} $:
\begin{equation}
    e(x) \; := \; \bar{D} E(x) \; = \; \frac{- x}{ \sigma_{n} |x|^{n} }
\end{equation}

Given  $ G \subset {\mathbb{R}}^{n}  $  domain with Liapunov
boundary $ S := \partial G $ and  $u \in C^{1} ( G , Cl_{0,n} )
\cap C( \bar{G},Cl_{0,n} ) $:
\begin{definition}
\end{definition}
CAUCHY-BITSADZE OPERATOR:
\begin{equation}
    ( F_{S} u) ( x) \; := \; \int_{S} e (x-y) u(y) n_{S}(y) dS_{y}
\end{equation}
one has the following \cite{Gurlebeck-Sprossig-97}:

\begin{theorem} \label{th:Gurlebeck-Sprossig's (0,n)-Cauchy integral formula}
\end{theorem}
GURLEBECK-SPROSSIG'S $(0,n)$-CAUCHY INTEGRAL FORMULA:

\begin{hypothesis}
\end{hypothesis}
\begin{equation*}
     G \subset {\mathbb{R}}^{n}  \text{ domain with Liapunov
boundary }  S := \partial G
\end{equation*}
\begin{equation*}
    u \in Ker D
\end{equation*}
\begin{thesis}
\end{thesis}
\begin{equation*}
   ( F_{S} u ) (x) \; = \; \left\{%
\begin{array}{ll}
    u(x), & \hbox{if $ x \in G $;} \\
    0, & \hbox{if $ x \in {\mathbb{R}}^{n} - \bar{G}$.} \\
\end{array}%
\right.
\end{equation*}
\begin{remark}
\end{remark}
 Let us observe that theorem\ref{th:Gurlebeck-Sprossig's (0,n)-Cauchy integral
 formula}, requiring that $ n > 2 $, doesn't allow to recover the
 complex Cauchy integral formula as the $(0,1)$-case. So it is not
 so clear, at least to us, in which sense theorem\ref{th:Gurlebeck-Sprossig's (0,n)-Cauchy integral
 formula} is a generalization of Cauchy' integral formula.

 Let us observe, furthermore, that not contemplating the
 $(n,0)$-cases, such a theorem doesn't allow to recover an
 hyperbolic Cauchy's integral formula as the $ (1,0)$-case.

 \smallskip

 A more advanced generalization of Cauchy's integral formula has
 been presented in \cite{Hestenes-Sobczyk-87}, \cite{Sobczyk-96b}.

 Since the axiomatization of Clifford algebras therein introduced
 differs from that presented in section  \ref{sec:The hyperbolic algebra as a bidimensional
Clifford algebra} we will briefly discuss their interrelations.

\begin{definition} \label{def:geometric algebra}
\end{definition}
GEOMETRIC ALGEBRA:

a set $  {\mathcal{G}} $,whose elements are called
\emph{multivectors}
\begin{itemize}
    \item endowed with two binary internal operations, the sum  and the
multiplication (called \emph{geometric product}), such that:
\begin{enumerate}
    \item the addition is commutative:
\begin{equation*}
    A + B \; = \; B + A \; \; \forall A,B \in  {\mathcal{G}}
\end{equation*}
    \item the addition and the multiplication are associative:
\begin{equation*}
    ( A + B) + C \; = \; A + ( B + C) \; \; \forall A,B,C \in  {\mathcal{G}}
\end{equation*}
\begin{equation*}
    ( A  B)  C \; = \; A  ( B C) \; \; \forall A,B,C \in  {\mathcal{G}}
\end{equation*}
    \item the multiplication is distributive w.r.t. addition:
\begin{equation*}
    A ( B + C ) \; = \; A B + A C \; \; \forall A,B,C \in  {\mathcal{G}}
\end{equation*}
\begin{equation*}
    (B+C) A \; = \; ( B A + C A ) \; \; \forall A,B,C \in  {\mathcal{G}}
\end{equation*}
    \item existence and uniqueness of additive and multiplicative
    identities:
\begin{equation*}
    \exists ! 0 \in  {\mathcal{G}} \; : \; A + 0 = A \; \; \forall A \in  {\mathcal{G}}
\end{equation*}
\begin{equation*}
    \exists ! 1 \in  {\mathcal{G}} \; : \; 1 A \; = \; A  \; \; \forall A \in  {\mathcal{G}}
\end{equation*}
    \item existence and uniqueness of the additive inverse:
\begin{equation*}
   \forall A \in {\mathcal{G}} , \exists ! - A \in  {\mathcal{G}} \, : \,   A + (- A) \; = \; 0
\end{equation*}
    \item grade-decomposition of a multivector $ A \in
    {\mathcal{G}} $
\begin{equation*}
    A \; = \; \sum_{r \in {\mathbb{N}}} < A >_{r}
\end{equation*}
where $ < A >_{r} $ is called the \emph{r-vector part} of A and
where the \emph{r-grade operator} $ < \cdot >_{r} : {\mathcal{G}}
\mapsto {\mathcal{G}}^{r} $ ( $  {\mathcal{G}}^{r} $ being the
subalgebra of $ {\mathcal{G}} $ formed by   the \emph{r-vectors},
i.e. the set of the multivectors such that $ A = < A >_{r}$, with
the assumption that $   {\mathcal{G}}^{0} = {\mathbb{R}} $ ), is
such that:
\begin{equation*}
    < A + B >_{r} \; = \; < A >_{r} + < B >_{r} \; \; \forall A ,
    B \in  {\mathcal{G}} ,  \forall r \in {\mathbb{N}}
\end{equation*}
\begin{equation*}
    < \lambda A >_{r} \; = \; \lambda < A >_{r} \; = \; < A >_{r}
    \lambda \; \; \forall A \in  {\mathcal{G}} , \forall \lambda
    \in {\mathcal{G}}^{0} , \forall r \in {\mathbb{N}}
\end{equation*}
\begin{equation*}
  < < A >_{r} >_{r} \; = \; < A >_{r} \; \; \forall A \in
  {\mathcal{G}} , \forall r \in {\mathbb{N}}
\end{equation*}
    \item the pseudo-euclidean condition:
\begin{equation*}
    a^{2} \; := \;  a a \; = \; < a^{2} >_{0} \; \; \forall a \in {\mathcal{G}}^{1}
\end{equation*}
     \item simple r vectors
\begin{equation*}
    \forall A \in {\mathcal{G}}^{r}_{S} : A \neq 0 \; \exists a \in
    {\mathcal{G}}^{1} : a \neq 0  \; and \; A a \in {\mathcal{G}}^{r+1}_{S}
\end{equation*}
where the set $ {\mathcal{G}}^{r}_{S} $ of the \emph{simple
r-vectors} is defined as the set of the \emph{r-vectors} that can
be expressed as the product of r mutually anticommuting
\emph{1-vectors}
\end{enumerate}
    \item endowed with a \emph{reversion operator} $ \cdot^{\dag} : {\mathcal{G}} \mapsto {\mathcal{G}} $  such that:
\begin{equation*}
     (A B)^{\dag} \; = \; B^{\dag} A^{\dag} \; \; \forall A,B \in {\mathcal{G}}
\end{equation*}
\begin{equation*}
    ( A + B)^{\dag} \; = \; A^{\dag} +  B^{\dag} \; \; \forall A,B \in {\mathcal{G}}
\end{equation*}
\begin{equation*}
    <  A^{\dag} >_{0} \; = \;  <  A  >_{0} \; \; \forall A \in {\mathcal{G}}
\end{equation*}
\begin{equation*}
 a^{\dag} \; = \; a \; \; \forall a \in {\mathcal{G}}^{1}
\end{equation*}
\end{itemize}

\smallskip

The name \emph{reversion operator} is justified by the following:
\begin{theorem} \label{th:on the name of the reversion operator}
\end{theorem}
\begin{equation*}
    ( a_{1} \cdots a_{n})^{\dag} \; = \; a_{n} \cdots a_{1} \; \;
    \forall a_{1}, \cdots ,  a_{n} \in {\mathcal{G}}^{1}
\end{equation*}

\smallskip

Given $ A \in {\mathcal{G}}^{r} $ and $ B \in {\mathcal{G}}^{s} $:
\begin{definition}
\end{definition}
OUTER PRODUCT OF A AND B:
\begin{equation*}
    A \wedge B \; := \; < A B >_{r+s}
\end{equation*}
The outer product of two arbitrary multivectors $ A ,B \in
{\mathcal{G}} $ may be then defined as:
\begin{definition}
\end{definition}
OUTER PRODUCT OF A AND B:
\begin{equation*}
 A \wedge B \; := \; \sum_{r \in {\mathbb{N}}} \sum_{s \in
 {\mathbb{N}}} < A >_{r} \wedge < B >_{s}
\end{equation*}
Given $ A,B \in {\mathcal{G}}$:
\begin{definition} \label{def:scalar product of two multivectors}
\end{definition}
SCALAR PRODUCT OF A AND B:
\begin{equation*}
    A \star B \; := \; < A B >_{0}
\end{equation*}
One has that:
\begin{theorem} \label{th:special decomposition of a simple n-vector}
\end{theorem}

\begin{hypothesis}
\end{hypothesis}
\begin{equation*}
    A \in {\mathcal{G}}^{n}_{S}
\end{equation*}

\begin{thesis}
\end{thesis}
\begin{equation*}
    \exists a_{1} , \cdots , a_{n} \in {\mathcal{G}}^{1} : A =
    a_{1} \cdots a_{n} \: and \: A^{\dag} \star A = a_{1}^{2}
    \cdots a_{n}^{2}
\end{equation*}
Theorem\ref{th:special decomposition of a simple n-vector} allows
to introduce the following:
\begin{definition}
\end{definition}
$ A \in {\mathcal{G}}^{n}_{S} $ HAS SIGNATURE (p,r) :
\begin{equation}
    A =
    a_{1} \cdots a_{n} \: and \: A^{\dag} \star A = a_{1}^{2}
    \cdots a_{n}^{2} \: and \: card( \{ a_{i} : a_{i}^{2} > 0 \} )
    = p \: and \: card( \{ a_{i} : a_{i}^{2} < 0 \} ) = r
\end{equation}
We will denote the set of all the simple vectors with signature
(p,r) by $  {\mathcal{G}}^{(p,r)}_{S} $.

Given $ A \in  {\mathcal{G}}^{(p,r)}_{S} $:
\begin{definition}
\end{definition}
\begin{equation*}
 {\mathcal{A}}_{p,r} \; := \; \{ a \in {\mathcal{G}}^{1} \, : \, a  \wedge A \, = \, 0 \}
\end{equation*}
\begin{definition}
\end{definition}
\begin{equation*}
 {\mathcal{G}} ( {\mathcal{A}}_{p,r}   ) \; := \;  (
 {\mathcal{G}} , + , \text{ geometric product } , \cdot^{\dag} ) |_{{\mathcal{A}}_{p,r}}
\end{equation*}
The link with definition\ref{def:Clifford algebra with signature
(p,r)} is then given by the following:
\begin{theorem}
\end{theorem}
\begin{equation*}
 {\mathcal{G}} ( {\mathcal{A}}_{p,r}   ) \; = \; Cl_{(p,r)} \; \;
 \forall p,r  \in {\mathbb{N}}
\end{equation*}
\begin{corollary}
\end{corollary}
\begin{equation*}
  {\mathcal{G}} ( {\mathcal{A}}_{1,0} ) \; = \; {\mathbb{G}}
\end{equation*}

\smallskip

\begin{remark}
\end{remark}
ON THE METRIC STRUCTURE OF THE GEOMETRIC ALGEBRA

 Let us suppose to replace the \emph{pseudo-euclidean condition} in definition\ref{def:geometric algebra}
 with the more restrict \emph{euclidean condition}:
\begin{equation*}
    a^{2} \; := \;  a a \; = \; < a^{2} >_{0} \in (0,\infty) \; \;
    \forall a \in {\mathcal{G}}^{1} : a \neq 0
\end{equation*}
Is then possible to introduce the following:
\begin{definition}
\end{definition}
MAGNITUDE OF $ a \in {\mathcal{G}}^{1} $
\begin{equation*}
    | a | \; := \; \sqrt{a^{2}}
\end{equation*}
Theorem \ref{th:on the name of the reversion operator} can then be
used to infer that:
\begin{theorem}
\end{theorem}
\begin{equation*}
    A^{\dag} \star A  \; \geq \; 0 \; \; \forall A \in  {\mathcal{G}}
\end{equation*}
\begin{proof}
Given:
\begin{equation}
    A \; = \; \prod_{i=1}^{n} a_{i} \; : \; a_{i} a_{j} = - a_{j}
    a_{i} \; \; \forall i \neq j
\end{equation}
one has that:
\begin{equation}
    ( a_{1} \cdots a_{n} )^{\dag} \star ( a_{1} \cdots a_{n} ) \;
    = \; ( a_{n} \cdots a_{1} ) \star ( a_{1} \cdots a_{n} ) \; =
    \prod_{i=1}^{n} | a_{i} |^{2} \geq 0
\end{equation}
So, introduced the notation:
\begin{equation}
   {\mathcal{G}}_{S} \; := \; \cup_{n \in {\mathbb{N}}} {\mathcal{G}}_{S}^{n}
\end{equation}
we have proved that:
\begin{equation*}
    A^{\dag} \star A  \; \geq \; 0 \; \; \forall A \in  {\mathcal{G}}_{S}
\end{equation*}
The case of non-simple multivectors can then be reduced to that of
simple multivectors to get the thesis
\end{proof}

It is then possible to introduce the following:
\begin{definition} \label{def:magnitude of a multivector in the euclidean case}
\end{definition}
MAGNITUDE OF $ A \in {\mathcal{G}} $
\begin{equation*}
    | A | \; := \; \sqrt{A^{\dag} \star A}
\end{equation*}

and the induced distance:
\begin{definition} \label{def:metric over the geometric algebra}
\end{definition}
$ d : {\mathcal{G}} \times {\mathcal{G}} \mapsto [0, + \infty)$
\begin{equation*}
    d ( A,B ) \; := \;  | A - B |
\end{equation*}
endowed with which the geometric algebra is a metric space on
which limits may be defined in the usual way \cite{Reed-Simon-80}.

Contrary, if, as we did in definition\ref{def:geometric algebra},
one doesn't assume the \emph{euclidean condition} but only the
\emph{pseudo-euclidean condition}, $ a^{2} = < a^{2} >_{0} a \in
{\mathcal{G}}^{1}$ and hence $ A^{\dag} \star A \; A \in
{\mathcal{G}} $ may become negative.

In this case definition \ref{def:magnitude of a multivector in the
euclidean case} has to be replaced with the following:
\begin{definition} \label{def:magnitude of a multivector in the pseudo-euclidean case}
\end{definition}
MAGNITUDE OF $ A \in {\mathcal{G}} $
\begin{equation*}
    | A | \; := \; \sqrt{ | A^{\dag} \star A |}
\end{equation*}

\smallskip

\begin{theorem}
\end{theorem}
\begin{hypothesis}
\end{hypothesis}
\begin{equation*}
    {\mathcal{A}}_{n} \subset  {\mathcal{G}} \text{ n-dimensional linear subspace}
\end{equation*}
\begin{thesis}
\end{thesis}
\begin{equation*}
  \exists ! (+I,-I) I \in {\mathcal{G}}_{S}^{n} \, : \,  \forall a_{1} , \cdots , a_{n} \in {\mathcal{G}}^{1} \cap {\mathcal{A}}_{n} , \exists
    \lambda \in {\mathcal{G}}^{0} \; : \\
      | I | = 1   \, and \, a_{1} \wedge  \cdots \wedge a_{n} \, = \, \lambda I
\end{equation*}
Given $ {\mathcal{A}}_{n} \subset  {\mathcal{G}} \text{
n-dimensional linear subspace}$:
\begin{definition}
\end{definition}
PSEUDOSCALARS OF $ {\mathcal{A}}_{n} $:
\begin{equation*}
    PS( {\mathcal{A}}_{n} ) \; := \; \{ \lambda I , \lambda \in {\mathcal{G}}^{0} : \lambda \neq 0    \}
\end{equation*}
\begin{theorem}
\end{theorem}
\begin{hypothesis}
\end{hypothesis}
\begin{equation*}
    {\mathcal{A}}_{n} \subset  {\mathcal{G}} \text{ n-dimensional linear subspace}
\end{equation*}
\begin{equation*}
a_{1} , \cdots , a_{n} \in {\mathcal{G}}^{1} \cap
{\mathcal{A}}_{n}
\end{equation*}
\begin{thesis}
\end{thesis}
\begin{equation*}
    a_{1} \wedge \cdots \wedge a_{n} \in  PS( {\mathcal{A}}_{n} )
    \; \Leftrightarrow \; a_{1} , \cdots , a_{n} \text{ linearly independent}
\end{equation*}

\medskip

Given a set $ {\mathcal{M}} \subset {\mathcal{G}}^{1}  $, let us
demand to \cite{Hestenes-Sobczyk-87} and \cite{Doran-Lasenby-03}
as to the determination of the conditions under which $
{\mathcal{M}} $ is said to be a vector manifold.

Given a vector manifold $  {\mathcal{M}} $, a point $ x \in
{\mathcal{M}} $  and a  1-vector $ a(x) \in {\mathcal{G}}^{1} $:
\begin{definition}
\end{definition}
a(x) IS TANGENT TO  $  {\mathcal{M}} $ in x:
\begin{equation*}
    \exists C : [0 , 1] \mapsto {\mathcal{M}} \; curve \; : C(0) =
    x \; and \; \frac{ d C ( \tau )  }{ d \tau }|_{\tau=0}
    = a(x)
\end{equation*}
We will denote the set of all the tangent vectors to  $
{\mathcal{M}} $ in x by $ {\mathcal{A}}(x) $.

Let us assume that  $ {\mathcal{A}}(x) $ is nonsingular, i.e. that
it posseses a unit pseudoscalar I(x) which we will call \emph{the
unit pseudo-scalar of  $ {\mathcal{M}} $ in x }.

\begin{definition}
\end{definition}
$ {\mathcal{M}}$ IS CONTINUOUS:
\begin{equation*}
    I(x) \text{ continuous in x} \; \; \forall x \in  {\mathcal{M}}
\end{equation*}
\begin{definition}
\end{definition}
ORIENTATION ON $ {\mathcal{M}} $

the assigment on $ {\mathcal{M}} $ of a continuous pseudoscalar.

\smallskip

\begin{definition}
\end{definition}
$ {\mathcal{M}} $ IS ORIENTABLE

its unit pseudoscalar I(x) is single valued

\smallskip

\begin{definition}
\end{definition}
$ {\mathcal{M}} $ IS SMOOTH

its unit pseudoscalar I(x) has derivative of all orders $ \forall
x \in {\mathcal{M}} $

\smallskip

Given an m-dimensional smooth oriented vector manifold $
{\mathcal{M}} $ an  a map $ f :  {\mathcal{M}} \mapsto
{\mathcal{G}}$:
\begin{definition}
\end{definition}
DIRECTED INTEGRAL OF f OVER $  {\mathcal{M}} $:
\begin{equation*}
    \int_{{\mathcal{M}}} d X \, f \; := \; \lim_{n \rightarrow
    \infty} \sum_{i=1}^{n} \Delta X(x_{i}) f( x_{i})
\end{equation*}
where the limit on the right side is to be understood in the usual
sense of Riemann integration theory.

\begin{definition} \label{Hestenes-Sobczyk derivative}
\end{definition}
HESTENES-SOBCZYK DERIVATIVE OF f:
\begin{equation*}
    \partial f(x) \; := \; \lim_{ | {\mathcal{R} |} \mapsto 0 } \frac{I^{-1}(x)}{| {\mathcal{R}}
    |(x)} \oint d S f
\end{equation*}
where:
\begin{enumerate}
    \item $ {\mathcal{R}} $ is an open smooth m-dimensional
    submanifold of  $ {\mathcal{M}} $ with x as an interior point
    \item the directed integral of f is taken over the boundary of
    $ {\mathcal{R}} $. The $(dim {\mathcal{M}}-1)$-vector dS representing a
    directed volume element of $ \partial {\mathcal{R}} $ is
    oriented so that:
\begin{equation}
    dS( x') n(x') \; = \; I( x') | dS( x') |
\end{equation}
where $ n(x')$ is the outward unit normal vector at a point $ x'
\in \partial {\mathcal{R}} $
 \item the limit is taken by shrinking $ {\mathcal{R}}$ and hence
 its volume  $ | {\mathcal{R}} |$ to zero at the point x; we allow
 the limit to be proportional to a delta-function or its
 derivatives, so that it is well defined in the sense of
 distribution theory \cite{Reed-Simon-80}.
\end{enumerate}

\begin{definition}
\end{definition}
f IS HESTENES-SOBCZYK ANALYTIC ON $ {\mathcal{M}} $:
\begin{equation*}
    \partial f(x) \; = \; 0 \; \; \forall x \in {\mathcal{M}}
\end{equation*}

Let us now consider the following:
\begin{definition}
\end{definition}
GREEN FUNCTION OF THE HESTENES-SOBCZYK DERIVATIVE

$ g : ({\mathcal{M}} \cup \partial  {\mathcal{M}} ) \times
({\mathcal{M}} \cup \partial  {\mathcal{M}} ) \mapsto
{\mathcal{G}}$:
\begin{equation}
 \partial g \; = \; \delta
\end{equation}
where the Dirac $ \delta $ distribution is defined by:
\begin{equation}
    \int_{{\mathcal{R}}} | d X( x' ) | \delta ( x - x') 'F( x') \;
    = \; F(x)
\end{equation}
with $   {\mathcal{R}} $ being a subregion of  $ {\mathcal{M}}$
containing x and with  $ F : {\mathcal{R}} \mapsto  {\mathcal{G}}
$ continuous.

We have at last all the necessary notions to present the
following:
\begin{theorem} \label{th:Hestenes-Sobczyk's generalized Cauchy integral formula}
\end{theorem}
HESTENES-SOBCZYK'S GENERALIZED CAUCHY INTEGRAL FORMULA:

\begin{hypothesis}
\end{hypothesis}
\begin{center}
 $ f :  {\mathcal{M}} \mapsto {\mathcal{G}}$ analytic
\end{center}
\begin{thesis}
\end{thesis}
\begin{equation*}
    f(x) \; = \; \frac{ (-1)^{dim {\mathcal{M}} } }{I(x)} \oint_{\partial {\mathcal{M}}}
     g(x,x') f(x') dS ( x') \; \; \forall x \in {\mathcal{M}}
\end{equation*}

\smallskip

Let us now analyze what Theorem\ref{th:Hestenes-Sobczyk's
generalized Cauchy integral formula} tells us as to the particular
case $  {\mathcal{M}} := {\mathcal{G}} (  {\mathcal{A}}_{1,0}) =
{\mathbb{G}} $ and $ f: {\mathcal{M}} \mapsto  {\mathcal{M}} $. At
this purpose it is sufficient to follow step  by step the analysis
concerning the geometric algebra of the plane ${\mathcal{G}}_{2}$
performed in \cite{Doran-Lasenby-03} adapting it to the case of
the bidimensional Minkowski spacetime algebra, i.e. the
 four-dimensional Clifford algebra $ {\mathcal{G}}_{2}^{hyp} $
spanned by the basis set consisting in:
\begin{enumerate}
    \item 1 one scalar
    \item $ \{ e_{0} , e_{1} \} $ two 1-vector
    \item $ I := e_{0} \wedge e_{1} $ one 2-vector
\end{enumerate}
where $ e_{0} $ and $ e_{1} $ are 1-vectors such that:
\begin{equation}
    e_{0}^{2} \; = \; - 1
\end{equation}
\begin{equation}
    e_{1}^{2} \; = \; 1
\end{equation}
\begin{equation} \label{eq:basis 1-vectors are orthogonal}
    e_{0} \cdot  e_{1} \; = \; 0
\end{equation}
One has that:
\begin{equation}
    e_{0} e_{1} \; = \;  e_{0} \cdot  e_{1} + e_{0} \wedge e_{1}
    \; = \;  e_{0} \wedge e_{1} \; = \; - e_{1} \wedge e_{0}
\end{equation}
i.e. $ e_{0} $ and $ e_{1} $ anticommute.

Let us observe furthermore that:
\begin{equation} \label{the square of the bivector}
 I^{2} \; = \; e_{0} e_{1} e_{0} e_{1} \; = \; - e_{0}^{2}
 e_{1}^{2} \; = \; + 1
\end{equation}
so that:
\begin{equation}
    {\mathbb{G}} \; = \; \{x + I y \, \, x,y \in {\mathbb{R}} \}
    \; \subset \;  {\mathcal{G}}_{2}^{hyp}
\end{equation}
One has that:
\begin{equation} \label{eq:first flip}
    I e_{0} \; = \; e_{0} e_{1} e_{0} \; = \; -  e_{0}^{2} e_{1}
    \; = \; e_{1}
\end{equation}
\begin{equation}
    I e_{1} \; = \; e_{0} e_{1} e_{1} \; = \; e_{0} e_{1}^{2} \; =
    \; e_{0}
\end{equation}
\begin{equation}
    e_{0} I \; = \; e_{0} e_{0} e_{1} \; = \; e_{0}^{2} e_{1} \; =
    \; - e_{1}
\end{equation}
\begin{equation}
    e_{1} I \; = \; e_{1} e_{0} e_{1} \; = \; - e_{1}^{2} e_{0} \;
    = \; - e_{0}
\end{equation}
from which it follows that I anticommutes with $ e_{0} $ and $
e_{1} $.

Observing preliminarily that:
\begin{equation}
    {\mathbb{R}}^{2} \; = \; \{ x e_{0} + y e_{1} , x,y \in {\mathbb{R}}
    \} \; \subset \;  {\mathcal{G}}_{2}^{hyp}
\end{equation}
let us introduce the following map $ F : {\mathbb{G}} \rightarrow
{\mathbb{R}}^{2}$:
\begin{equation}
    F ( z ) \; := z e_{0}
\end{equation}
One has that:
\begin{theorem}
\end{theorem}
\begin{equation}
    F( x + I y) \; = \; x e_{0} + y e_{1} \; \; \forall x,y \in {\mathbb{R}}
\end{equation}
\begin{proof}
\begin{equation}
    ( x+ I y) e_{0} \; = \; x e_{0} + y I e_{0}
\end{equation}
The thesis follows by eq.\ref{eq:first flip}
\end{proof}

\begin{corollary}
\end{corollary}
\begin{equation}
    F^{-1} ( x e_{0} + y e_{1} ) \; = \; x + I y \; = \; - ( x e_{0} + y e_{1} ) e_{0}  \; \; \forall x,y \in {\mathbb{R}}
    e_{0}
\end{equation}
Let us now introduce the following functional $ {\mathcal{F}}:
MAP( {\mathbb{G}}, {\mathbb{G}}) \, \mapsto \,
MAP({\mathbb{R}}^{2}, {\mathbb{R}}^{2})  $:
\begin{equation}
 {\mathcal{F}} [ \psi ] \; := \; F \circ \psi \circ F^{- 1}
\end{equation}
One has that:
\begin{theorem} \label{th:equivalence among analicity and Hestenes-Sobczyk analicity}
\end{theorem}

\begin{hypothesis}
\end{hypothesis}
\begin{equation*}
    f = u + I v \in MAP( {\mathbb{G}}, {\mathbb{G}})
\end{equation*}
\begin{thesis}
\end{thesis}
\begin{equation*}
    f \text{ is analytic} \; \Leftrightarrow \;  {\mathcal{F}} [ f ] \text{ is Hestenes-Sobczyk analytic}
\end{equation*}
\begin{proof}
 Let us observe, first of all, that the Hestenes-Sobczyk derivative of
definition\ref{Hestenes-Sobczyk derivative} reduces in our case
to:
\begin{equation}
    \nabla \; := \: e_{0} \frac{ \partial  }{ \partial x } + e_{1}
     \frac{ \partial  }{ \partial y }
\end{equation}
One has that:
\begin{equation}
 \nabla   {\mathcal{F}} [f ] \; = \; ( e_{0} \partial_{x} + e_{1}
 \partial_{y}) ( u e_{0} + v e_{1} ) \; = \; - \partial_{x} u +
 \partial_{y} v+ I ( \partial_{x} v - \partial_{y} u)
\end{equation}
and hence:
\begin{equation}
    \nabla f = 0 \; \Leftrightarrow \;   \partial_{x} u =
 \partial_{y} v   \; and \; \partial_{x} v = \partial_{y} u
\end{equation}
By theorem\ref{th:Cauchy-Riemann conditions in the hyperbolic
plane} the thesis immediately follows
\end{proof}

Let us now observe that the key point of the Cauchy Integral
Formula of theorem\ref{th:complex Cauchy integral formula} ,
expressed in terms of the geometric algebra $ {\mathcal{G}}_{2}
$,is that the Cauchy kernel $ \frac{1}{z'-z'_{0}}$ is the Green
function of the Hestenes-Sobczyk derivative:
\begin{equation}
  {\mathcal{F}}' [  \frac{1}{z'-z'_{0}} ] \; = \; \frac{r'-r'_{0}}{(r'-r'_{0})^{2} }
\end{equation}
\begin{equation}
    \nabla \frac{r'-r'_{0}}{(r'-r'_{0})^{2} } \; = \; 2 \pi \delta ( r'
    - r'_{0})
\end{equation}
where the basis $ \{ 1 ,  e_{0}' ,  e_{1}' , I':= e_{0}' \wedge
e_{1}' \} $ spanning   $  {\mathcal{G}}_{2} $ is defined by the
conditions:
\begin{equation}
    e_{0}'^{2} \; = \; e_{1}'^{2} \; = \; 1 \; \; and  \; \;
    e_{0}' \cdot  e_{1}' = 0
\end{equation}
where:
\begin{equation}
    {\mathbb{C}} \; = \; \{ x + I' y , x,y \in
    {\mathbb{R}} \} \; \subset \; {\mathcal{G}}_{2}
\end{equation}
 where:
\begin{equation}
    r' \; := \; x  e_{0}' +y e_{1}'
\end{equation}
\begin{equation}
    r'_{0} \; = \; x_{0}  e_{0}' + y_{0} e_{1}'
\end{equation}
and where $  {\mathcal{F}}' : MAP ( {\mathbb{C}}, {\mathbb{C}} )
\mapsto MAP ( {\mathbb{R}}^{2} , {\mathbb{R}}^{2}) $ is such that:
\begin{equation}
 {\mathcal{F}}' [ \psi ] \; := \;  F' \circ \psi \circ F'^{- 1}
\end{equation}
with $ F' : {\mathbb{C}} \mapsto {\mathbb{R}}^{2} $ such that:
\begin{equation}
  F' ( x + I' y) \; := \;  x  e_{0}' +y e_{1}'
\end{equation}

 In the case of $  {\mathcal{G}}_{2}^{hyp} $, contrary, one has
that:
\begin{equation}
    {\mathcal{F}} [ \frac{1}{ z - z_{0} } ] \; = \; \frac{( x- x_{0}) e_{0} - ( y - y_{0}) e_{1} }{( x - x_{0})^{2} - ( y - y_{0})^{2} }
\end{equation}
and hence:
\begin{equation}
    \nabla  {\mathcal{F}} [ \frac{1}{ z - z_{0} } ] \; \neq \; 2
    \pi \delta ( r - r_{0} )
\end{equation}
In analogy with what we saw for the complex case, it is natural,
anyway, to suppose that an Hyperbolic Cauchy Integral formula may
be obtained by determining the Green's function of the $
{\mathcal{G}}_{2}^{hyp} $'s Hestenes-Sobczyk derivative.

Such a subject is under investigation.
\newpage
\section{Acknoledgements}
We would like to thank prof. Garret Sobczyk for some useful
suggestions.

Our work was funded by the EU Training Network on "Quantum
Probability and Applications in Physics, Information Theory and
Biology".

\end{document}